\documentclass[a4paper,fleqn]{cas-sc}
\usepackage{algorithm}
\usepackage{algpseudocode}
\usepackage{amsmath, amssymb} 
\usepackage{graphicx}
\usepackage{array}
\usepackage{caption}
\usepackage{multirow}
\usepackage{booktabs} 
\usepackage{float}
\usepackage[T1]{fontenc} 
\usepackage[utf8]{inputenc} 
\usepackage{multirow}
\usepackage{caption}
\usepackage{textcomp}
\usepackage{geometry}
\usepackage{hyperref}
\usepackage{booktabs}
\usepackage{amsmath}
\usepackage{bm}
\usepackage{natbib}
\usepackage{flushend}
\usepackage{xcolor}
\usepackage[nameinlink,capitalize]{cleveref}
\hypersetup{colorlinks=true, linkcolor=blue, filecolor=cyan, citecolor=blue, urlcolor=blue}
\usepackage[short labels]{enumitem}
\usepackage[english] {babel}
\addto\extrasenglish{

}

\def\tsc#1{\csdef{#1}{\textsc{\lowercase{#1}}\xspace}}
\tsc{WGM}
\tsc{QE}
\begin{document}
\let\WriteBookmarks\relax
\def\floatpagepagefraction{1}
\def\textpagefraction{.001}



 \shorttitle{ A Social Context-aware Graph-based Multimodal Attentive Learning Framework}
\shortauthors{Shahid \textit{et al.}}
\title [mode = title]{ A social context-aware graph-based multimodal attentive learning framework for disaster content classification during emergencies: A benchmark dataset and method
}                      

\author[1]{Shahid Shafi Dar}[auid=1]
\cormark[1]
\ead{phd2201201004@iiti.ac.in}
\cortext[cor1]{Corresponding author}

\author[1]{Mohammad Zia Ur Rehman}
\ead{phd2101201005@iiti.ac.in}

\author[2]{Karan Bais}
\ead{karanbais2701@gmail.com}

\author[3]{Mohammed Abdul Haseeb}
\ead{20bcs085@iiitdwd.ac.in}
\author[1]{Nagendra Kumar}
\ead{nagendra@iiti.ac.in}

\affiliation[1]{organization={Indian Institute of Technology Indore}, postcode={453552}, country=India}
\affiliation[2]{organization={Institute of Engineering and Technology, Devi Ahilya Vishwavidyalaya }, postcode={452017}, country=India}

\affiliation[3]{organization={Indian Institute of Information Technology Dharwad}, postcode={580009}, country={India}}

\maketitle
\begin{abstract}
In times of crisis, the prompt and precise classification of disaster-related information shared on social media platforms is of paramount importance for effective disaster response and public safety. During such critical events, people utilize social media as a medium for communication, sharing multimodal textual and visual content. However, due to the substantial influx of unfiltered and diverse data, humanitarian organizations face challenges in effectively leveraging this information. Numerous methods have been proposed for classifying disaster-related content, but these methods lack modeling users’ credibility, emotional context, and social interaction information, which is crucial for classification. In this context, we propose a method, CrisisSpot, that leverages a Graph-based Neural Network to comprehend intricate relationships between textual and visual modalities and Social Context Features to incorporate user-centric and content-centric information. We also propose Inverted Dual Embedded Attention, which captures both harmonious and contrary patterns present in the data to harness complex interactions and facilitate richer insights in multimodal data. We have developed a multimodal disaster dataset, TSEqD (Turkey-Syria Earthquake Dataset), which is a large annotated dataset for a single disaster event containing 10,352 data samples. Through extensive experimentation, CrisisSpot has demonstrated significant improvements, achieving an average gain of 9.45\%  and 5.01\% in F1-score compared to the state-of-the-art methods on the publicly available CrisisMMD dataset and TSEqD dataset, respectively.
\end{abstract}


\section{Introduction}
The rising frequency and growing severity of natural disasters, such as earthquakes, floods, and hurricanes, cause a significant socio-economic disturbance on a global scale~\citep{disaster}. These events have a devastating impact on the economy, human health, and livelihood. In 2022 alone, there were 387 natural disasters, resulting in 30,704 deaths and affecting approximately 185 million people, causing millions of dollars of damage costs as reported by the Emergency Event Database (EM-DAT)\footnote{https://reliefweb.int/report/world/2022-disasters-numbers}.  In addition to physical damages, these disasters often lead to homelessness, mental health issues, unemployment, emotional distress, financial difficulties, and economic recession. Therefore, it is crucial to consistently explore innovative and enhanced approaches for response and recovery during disasters.\\

\textit{\textbf{Role of Social Media during Disaster Events}} \\
Social media platforms have emerged as invaluable real-time information sources during disaster events \citep{8715653}. These platforms serve as channels for the rapid dissemination of information, providing critical real-time updates and firsthand accounts during such events \citep{10.1145/2771588, SocialMediaBenefits}. For instance, during the 2023 Turkey-Syria earthquake, social media\footnote{https://news.northeastern.edu/2023/02/08/social-media-turkey-earthquake/} served as an indispensable communication platform, facilitating rescue requests and streamlining the coordination of aid efforts.  
Numerous individuals who were stuck beneath debris utilized mobile tools to access social media platforms, seeking rescue assistance by sharing messages on social media\footnote{https://www.theguardian.com/world/video/2023/feb/07/turkey-earthquake-victim-rescued-after-social-media-plea-under-rubble-video}, which signifies the utility of social media during such events. However, the research community faces numerous challenges in utilizing social media information during crisis events.\\

\textit{\textbf{Challenges in Leveraging Social Media Information}}\\
The real-time information provided by social media platforms enhances crisis management capabilities \citep{SMD}, but the overwhelming influx of irrelevant data delays disaster response and management \citep{SocialMediabenfits} as manual segregation of the relevant information is a time-consuming and costly process. Identifying relevant content amidst this digital noise is a significant challenge. Another important aspect is the incorporation of a diverse mixture of multimodal data \citep{MMLSURVEY, multimodal}, encompassing images and text. Each of these modalities presents its own complexities. For instance, the text in crisis-related social media posts often uses informal language, contains frequent misspellings, incorporates numerous hashtags, and mentions other users. Relying solely on text analysis may not capture the full context of a situation \citep{zishan}. Conversely, images, when viewed in isolation, may lack the necessary context for effective decision-making. To this end, we propose CrisisSpot, a robust and versatile disaster content classification method. The proposed approach merges linguistic context with visual evidence to create a more comprehensive understanding of disaster situations.  A system like CrisisSpot empowers stakeholders by allowing them to effectively classify and respond to the most pertinent information contained within social media data posted during a disaster. \\

\textit{\textbf{Existing Multimodal Paradigms and Limitations}} \\
It is worth noting that previous works \citep{Previous_work1, Previous_work2,zishan, Correlation, Alam2018} have often overlooked the utilization of Social Context Features (SCF). SCF \citep{SCF} is an umbrella term that consists of social interaction metrics and semantic text information. They provide a multifaceted view of disaster-related content by incorporating factors such as the post-user history \citep{credibility}, emotional context~\citep{Emotion}, and user engagement.
In the existing works, various approaches have been proposed over time \citep{ multimodal1, zishan, Correlation, Alam2018, CABD, MANN}. \citep{Alam2018} utilize the K-nearest neighbor method for constructing a nearest-neighbor graph. In their methodology, nodes are updated based on intra-cluster information, while CrisisSpot updates nodes using neighborhood information selected from the entire dataset across multiple iterations. Consequently, CrisisSpot provides a comprehensive overview of data to each node for feature refinement. Additionally, their choice of neighborhood based on Euclidean distance between nodes is suboptimal for high-dimensional data spaces~\citep{cosine_similarity}.  To this end, CrisisSpot employs cosine similarity for neighbor selection, a metric proven to perform better in high-dimensional spaces. \\
\citep{CABD} utilize SSE-Graph~\citep{SSE}, which relies on the knowledge graph for swapping embeddings, but it may not explicitly capture the complex structural relationships within the graph. Also, depending on the size of the knowledge graph and the frequency of swapping, SSE-Graph may face scalability challenges. Handling large-scale graphs with frequent embedding swaps might lead to increased computational complexity. To this end, CrisisSpot excels at learning representations that capture the local graph topology, which can be crucial for certain tasks. Additionally, our graph-based approach is designed for scalable graph learning and may provide more efficient solutions for large graphs~\citep{GraphSage}.\\ 
The aforementioned studies~\citep{zishan, Correlation, CABD} employ conventional attention mechanisms such as cross-attention to focus solely on aligned information. In contrast, our proposed attention mechanism, Inverted Dual Embedded Attention (IDEA), adeptly directs attention to both aligned and misaligned information patterns across diverse modalities. This explicit recognition significantly augments the model's capability to effectively handle multimodal data. In conclusion, while existing multimodal paradigms have made significant strides, our study highlights some important research gaps. Our work addresses these limitations by integrating multimodal features, acknowledging the importance of Social Context Features (SCF), and introducing the Inverted Dual Embedded Attention (IDEA) mechanism.\\

\textit{\textbf{Addressing Research Gaps: Our Solution at a Glance}}\\
In response to these gaps in the existing literature, our work introduces a novel approach based on the multimodal attentive framework for disaster content classification. Our approach serves a dual purpose: first, we classify data into informative and non-informative content. Subsequently, we classify the informative content into various humanitarian categories. To achieve this, we employ a graph-based approach that leverages intricate relationships among disaster-related content, enriching the model's contextual understanding. This graphical representation significantly enhances our model's ability to discern subtle nuances within crisis-related information. Furthermore, we propose an attention mechanism tailored to identify both similar and dissimilar content pairs, thereby fostering a more nuanced comprehension of context. In addition to these advancements, our framework integrates Social Context Features, effectively utilizing text-centric and user-centric information. SCF, in our work, comprises User Informative Score (UIS), Crisis Informative Score (CIS), and User Engagement metrics. The UIS of SCF is of paramount importance in evaluating the user credibility of social media content. Similarly, the CIS of SCF measures the presence of crisis-related vocabulary in text, indicating a higher degree of informative content related to the crisis. It quantifies the semantic attributes of text. These features have often been neglected in previous works focused on disaster content classification. \\
In summary, our research presents a comprehensive and innovative approach to disaster content classification during emergencies. By harnessing the power of multimodal data, graph-based enriched context modeling, and Social Context Features, our framework equips emergency responders and humanitarian organizations with a more accurate and context-aware model. \\

\textit{\textbf{Key Contributions}}\\
The key contributions of this work  are outlined as follows:
\begin{enumerate}
    \item We propose a ``multimodal attentive learning framework'' for the problem of disaster content classification for various crisis events. The proposed method leverages graph-based and attention-based approaches for feature enrichment and more nuanced comprehension of the model's prediction ability.
    
    \item To the best of our knowledge, we are the first to leverage the potential of Social Context Features (SCF) for disaster content classification in multimodal settings. SCF, effectively merging text-centric and user-centric information, elevates model performance by considering content within a broader context encompassing user history, social interactions, and user post informativeness.

    \item We propose an attention mechanism, Inverted Dual Embedded Attention (IDEA), which captures both harmonious and contrary information present in the modalities. IDEA enables CrisisSpot to capture dependencies and complex relationships within the data, leading to more robust feature representations and richer comprehension of multimodal content.

    \item We have created a new dataset, TSEqD (Turkey-Syria Earthquake Dataset), which is a multimodal crisis dataset composed of 10,352 data samples. To the best of our knowledge, this is the largest single-event multimodal dataset annotated for both Informative and Humanitarian Tasks.
    
    \item  We have curated a list of 4,268 crisis-related lexicons, which are utilized to compute CIS, an important component of SCF. This comprehensive lexicon list may help researchers and humanitarian organizations efficiently assess and categorize disaster-related textual content.
    
    \item  We conduct comprehensive experiments on two datasets, such as CrisisMMD and TSEqD, to evaluate the effectiveness of our proposed approach.  The experimental outcomes clearly indicate that our approach surpasses state-of-the-art methods.

\end{enumerate}
The subsequent sections of this article are organized as follows: In~\autoref{Section.2}, we provide a comprehensive review of the relevant literature.\autoref{Section.3a} outlines our problem definition, ~\autoref{Section.3} outlines methodology, ~\autoref{Section.4} presents the outcomes of our experimental assessments.  ~\autoref{Section.5} presents discussions. Lastly, in ~\autoref{Section.6}, we provide our concluding remarks on this study.

\section{Related Work} \label{Section.2}
We examine previous research on disaster content classification, categorizing these studies into two primary groups: unimodal and multimodal disaster content classification.

\subsection{Unimodal Disaster Content Classification}
Unimodal disaster content classification involves the process of classifying disaster-related content using a single data modality, which includes text and images. Two prevalent instances of unimodal disaster content classification are disaster text classification and disaster image classification.

\subsubsection{Disaster Text Classification}
Most of the previous research~\citep{madichetty2021} on classifying disaster-related content has primarily concentrated on the textual modality. For instance, \citep{ghafarian2020} propose a method to identify informative tweets, leveraging the assumption that each tweet represents a word distribution. They perform experiments on 20 crisis incidents. \citep{rudra2018} employ vocabulary-agnostic features, including syntactic features and basic lexical elements, to classify tweets as either conveying situational or non-situational information. They conduct experiments using various disaster datasets and compare their approach with the Bag-Of-Words (BOW) model for Hindi and English tweets. \citep{kumar} evaluate multiple traditional Machine Learning and advanced Deep Learning algorithms for categorizing tweets related to disasters across six distinct classes. The effectiveness of the model is assessed through testing on four specific disaster scenarios, such as wildfires, earthquakes, hurricanes, and floods. \citep{xie} propose a supervised contrastive learning framework (SCMC) for multi-level text classification of disaster information in social media. This framework incorporates data reconstruction and an encoder-decoder architecture to effectively handle disaster-related information. They conduct experiments on three distinct datasets, demonstrating that their method leads to enhanced classification performance. \citep{constraint} propose a novel multi-label text classification approach that addresses the limitations arising from imbalanced label distribution and unclear label hierarchies. They introduce the multi-level constraint augmentation method that leverages historical generation information, sample text content, and topic to generate balanced and relevant samples, specifically targeting the issue of imbalanced labels. They also introduce a label association interaction mechanism that combines text information with label weights, allowing the model to capture co-occurrence patterns between labels and effectively distinguish similar labels. Nevertheless, prior approaches have made significant strides in addressing challenges associated with disaster text classification. However, these methods neglect crucial informative factors, such as user post-history, user engagement metrics, and the emotional context within the text. These factors play a vital role in enhancing the classification capabilities of a method, enabling it to identify relevant, reliable, and information-rich content within the text.

\subsubsection{Disaster Image Classification}
Due to the limited number of disaster image datasets, the predominant focus has shifted towards the development of comprehensive and high-quality image datasets \citep{incidents1m,alam2023medic} within the field of disaster management. Most of the research focuses on categorizing the images based on the severity of the damage. \citep{chaudhuri2020} collect geotagged images from regions impacted by earthquakes. They apply a deep learning method to categorize these images, with a specific focus on identifying survivors under the rubble. They find that the deep learning method exhibits improved accuracy in image classification compared to conventional machine learning techniques. \citep{nguyen2017} employ a deep CNN to classify disaster-related images into various categories, such as severe, mild, and no-damage classes. Their CNN-based approach surpassed Bag-of-Visual-Words (BoVW) methods, achieving significant F1-scores between 0.67 and 0.89. \citep{alam} conduct image classification of disaster-related images sourced from social media to assess the extent of the damage. Their work underscores the significance of domain-specific fine-tuning of deep Convolutional Neural Networks.
The existing methods have made substantial progress in tackling the issues related to classifying disaster images. However, they fail to address issues inherent to image data, such as variations in data quality and the context-dependent nature of interpretation. These elements significantly contribute to augmenting the model's capabilities in classifying images.

\subsection{Multimodal Disaster Content Classification}
In recent times, researchers have introduced multimodal systems that harness both text and images to extract relevant information. \citep{CrisisDIAS} use the CrisisMMD dataset \citep{CRISISMMD} to build a multimodal end-to-end gated system that identifies the infrastructural damage cues in the data and further qualifies the damage severity of the data based on the ordinal quantification given in the dataset. They surpass state-of-the-art methods by a significant amount. \citep{CABD}  propose a new cross-attention module during the data fusion process to avoid the propagation of false information. \citep{koshy2023multimodal} introduce an advanced Deep Learning framework that integrates various input modalities, incorporating text and images derived from user-generated content. Their approach outlines a systematic methodology for distinguishing informative tweets from the vast volume of user-generated content on social media platforms. \citep{madichetty2023roberta} have presented a novel technique for identifying informative tweets with multimodal content during disaster situations. Their method involves the utilization of pre-trained RoBERTa for text features and VGG-16 for image features. To combine these two model outputs, they employ a multiplicative fusion technique. Prior multimodal methods have made significant progress in handling the issues related to disaster content classification. However, most of the prior methods do not incorporate graph-based learning techniques, which helps us enrich the features. Additionally, they overlook various social interaction information, which enhances the classification abilities of the model.
 
\subsection{Attention Mechanisms}
Attention mechanisms are vital components of Deep Learning models, particularly for tasks involving sequential data such as text or image features. They enable models to focus on relevant parts of input data. Classical attention mechanisms include:
Additive Attention \citep{additive_attention}, Self Attention \citep{self}, and Cross Attention \citep{cross}. Additive attention utilizes a scoring function to assign weights to elements in a sequence, creating a weighted sum that highlights important information. They are widely used in tasks such as machine translation. Self-Attention focuses on analyzing the relationships within a single sequence and employs a scoring function to calculate attention scores. However, instead of comparing each element to a single vector, self-attention compares each element to all other elements in the sequence. This allows the model to identify long-range dependencies within the data, which can be crucial for tasks such as text summarization. Most of the existing works \citep{DMCC, RoBERTaMFT, CAMM, CABD} use Cross-Attention to analyze relationships between two sequences, useful in tasks such as machine translation or visual question answering. It allows models to leverage information from different modalities for better performance. However, these traditional attention mechanisms primarily focus on identifying supportive information within data. In complex domains such as disaster classification, sentiment analysis, and anomaly detection contrary information is also crucial. Current attention mechanisms struggle to distinguish between situations with aligned information and those with misaligned information. To tackle this, we introduce Inverted Dual Embedded Attention (IDEA). IDEA can identify both supportive and non-supportive information patterns within complex data. It achieves this with two modules: a Harmonious Attention Module (HAM) which focuses on elements reinforcing the main concept and a Contrary Attention Module (CAM) which attends to elements deviating from the main concept, capturing contrary information patterns. IDEA combines outputs from both modules to create a rich representation, enabling accurate differentiation between situations with clear information and those with potentially conflicting details, improving disaster classification.

\section{Problem Definition} \label{Section.3a}
Let $\mathcal{D} = \{\{T_i\},\{I_i\}\}_{i=1}^{N}$ be a multimodal dataset containing $N$  number of samples with $T_i = \{t_i\}_{i=1}^{N}$ represents the textual modality for the $i^\text{th}$ sample, where $t_i$ is the textual content and $I_i = \{i_i\}_{i=1}^{N}$ represents the visual modality for the $i^\text{th}$ sample and where, $i_i$ is the visual content. We are given a post containing text $t$ and image $i$, and our objective is to tackle the following tasks:\\\\
Task 1: Informative and  Non-Informative Content Classification \\
Task 1 is to classify the given post into informative and non-informative categories.
Let $y_i \in \{0, 1\}$ be the label for the $i^\text{th}$ post, where 0 represents non-informative and 1 represents the informative post. $f(t, i)$ denotes a function representing a model that takes text ($t$) and image ($i$) as inputs and produces a probability score indicating their informativeness. This task can be represented as shown in \autoref{Info}:
\begin{equation} \label{Info}
\hat{y}_i = \arg\max_{c \in C} P(y_i = c \mid t, i)
\end{equation}
where $\hat{y}_i$ is the predicted label for the $i^\text{th}$ post, $\arg\max_{c \in C}$ denotes finding the argument (class) that maximizes the probability and $C$ denotes the maximum number of classes, and $P(y_i = c | t, i)$ represents the conditional probability of the $i^{\text{th}}$ post belonging to class c given the text and image information.\\\\
Task 2: Humanitarian Content Classification\\
Task 2 is to classify the informative post into any seven critical and potentially actionable humanitarian categories.
Let $z_i \in \{1, 2, ..., 7\}$ be the subcategory label for the informative $i^\text{th}$ post, where $z_i$ can belong to any one of the categories: infrastructure and utility damage,  injured or dead people, rescue volunteering or donation effort, vehicle damage, affected individuals, missing or found people, and other relevant information.\\
We can define a similar function $g(t, i)$ that takes text and image information of an informative post and outputs a probability distribution over the seven categories as shown in \autoref{task2}:
\begin{equation} \label{task2}
P(z_i = k | t, i) \approx g_k(t, i)
\end{equation}
where $P(z_i = k | t, i)$ denotes the conditional probability of the informative $i^\text{th}$ post belonging to category $k$, $g_k(t, i)$ represents the predicted probability for category $k$ given text and image information obtained from the function $g$.\\
The research objectives are delineated as follows:
\begin{itemize}
\item Assessing the impact of Social Context Features on crisis predictions.
\item Investigating the impact of reconstructed features through Graph Neural Networks on the classification outcomes.
 \item To investigate whether the proposed attention mechanism, IDEA, captures aligned and misaligned information patterns.
\end{itemize}

\section{Methodology} \label{Section.3}
\begin{figure*}[pos=ht]
    \centering
    \includegraphics[width=\linewidth]{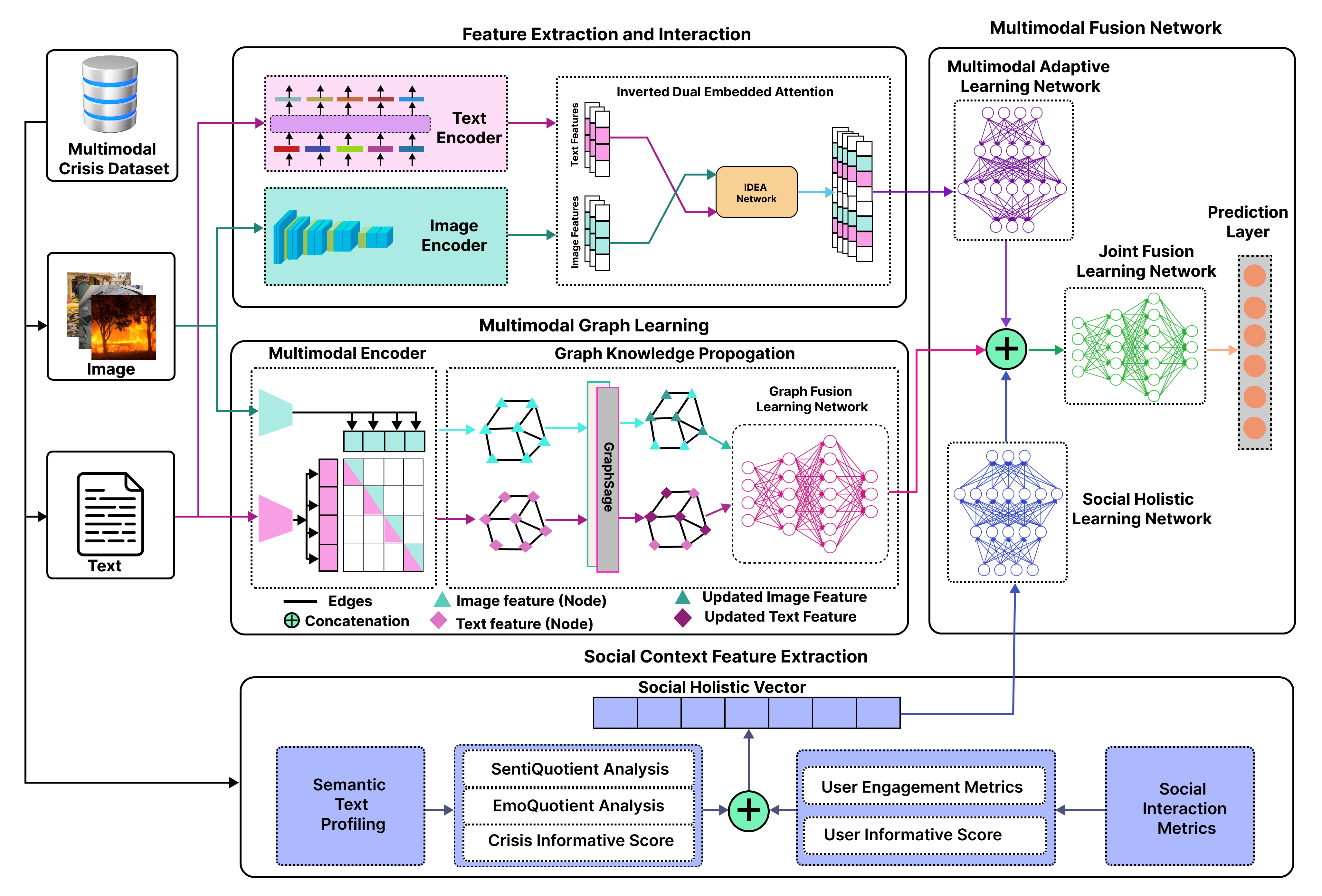}
    \caption{System Architecture of CrisisSpot.The features get updated after applying the GraphSAGE layer. The updated node features are depicted using a color representation that corresponds with darker shades.}
    \label{Fig.1}
\end{figure*}

We present our proposed method by presenting its comprehensive description along with the overall architectural framework, as depicted in \autoref{Fig.1}. The proposed methodology can be divided into four sub-modules, namely, Feature Extraction and Interaction, Multimodal Graph Learning, Social Context Feature Extraction, and Multimodal Fusion Network. Feature Extraction and Interaction focuses on extracting relevant features from text and visual modalities and capturing interactions via the multimodal attention mechanism (Inverted Dual Embedded Attention). Multimodal Graph Learning involves leveraging multimodal graph knowledge propagation to integrate information from text and image modalities. Social Context Feature Extraction concentrates on incorporating social interaction and text-based information into the model, enriching the model's context comprehension by incorporating valuable information from text-centric evaluations and user-oriented assessments.  Multimodal Fusion Network encompasses the fusion of multimodal representations of data from all the modules, yielding a comprehensive output followed by a prediction layer to generate crisis predictions.
\subsection{Feature Extraction and Interaction}
In this section, we present a discussion about feature extraction and interaction processes. This module comprises two submodules: (a) Textual and Visual Feature Extraction and (b)  Inverted Dual Embedded Attention.

\subsubsection{Textual Feature Extraction}
Each input modality is subjected to an initial encoding process, generating separate unimodal representations labeled as $H_t$ for the textual modality and $H_v$ for the visual modality.
In the context of text modality, given a sequence of text tokens denoted as $S_t= \{T_1, T_2, T_3, \ldots, T_n\}$, we employ the BERT-based~\citep{BERT} tokenization process to extract token embeddings, resulting in the matrix $H_t$ expressed in the following equation.
\begin{equation}
\label{eqn:1}
    H_t = {BERT}(S_t) \in \mathbb{R}^{d \times dt} 
\end{equation}
In \cref{eqn:1}, $H_t$ represents a matrix containing token embeddings, where $\mathbb{R}^{d \times dt}$ denotes a matrix with dimensions $(d, dt)$. We extract word-level features from the text encoder with a maximum sequence length of 128 and an embedding dimension of 768.
\subsubsection{Visual Feature Extraction}
The feature extraction process involves the utilization of the ResNet50 architecture, selectively extracting features up to the `$conv4\_block6\_3\_conv$' layer. The intention behind this decision is to capture complex and abstract representations that reside in deeper layers of the network. This is additionally performed to ensure that the shapes align with those of the text encoder, facilitating the interaction of features through the IDEA attention mechanism. The residual connections in ResNet50 facilitate effective gradient flow during training, addressing the vanishing gradient problem. The resulting feature map ($F_{\text{emb}}$) encapsulates hierarchical information, ranging from fundamental attributes such as edges and textures to more complex features such as object components and scenes. A bespoke convolutional layer, denoted as `$custom\_conv$', is seamlessly integrated into the architecture. This custom convolutional layer introduces 1024 filters and employs a (4, 4) kernel, inducing a transformative effect on the output shape, resulting in dimensions (11, 11, 1024). Subsequently, a crucial reshaping operation is applied to the feature map, yielding a tensor with dimensions (1, 121, 1024). This reshaping is strategically aligned with the subsequent processing requirements, ensuring optimal compatibility with the changing tensor structure. Simultaneously, a padding operation is introduced to the tensor to retain spatial information and maintain dimensional consistency. This step results in a tensor shape of (1, 128, 1024), crucial for the seamless progression of information through the network architecture. To safeguard the knowledge encoded in the ResNet50 pre-trained weights and fine-tune the model for the specialized image processing task, a discerning decision is made to freeze all original layers. This strategic freezing confines the training scope exclusively to the parameters of the `$custom\_conv$' layer. Equation (\ref{eqn:2}) details the calculation of each element in the final visual feature matrix $H_v$, where the variable \(i\) represents the index of the features in the final visual feature matrix ${H_v}$. $H_{\text{img}}$ and $W_{\text{img}}$ represent the height and width of the feature map, respectively, and $k$ denotes a specific channel within $F_{\text{emb}}$.
\begin{equation}
\label{eqn:2}
{(H_v)_i} = \frac{1}{H_{img} \cdot W_{img}} \sum_{w=1}^{W_{img}} \sum_{h=1}^{H_{img}} F_{{emb}}(h, w, k)
\end{equation}
The sequence of changes in the architecture, involving adjustments in feature extraction, custom convolutional layer integration, reshaping, and padding, results in a customized ResNet model. This tailored adaptation enhances the model's effectiveness for a specific image-processing application while leveraging the foundational knowledge inherent in the pre-trained ResNet-50 architecture.
\subsubsection{Inverted Dual Embedded Attention} \label{IDEA}
In multimodal data analysis, we often face the challenge of efficiently integrating information from various modalities, such as text and image, to achieve a holistic comprehension of the underlying content. Traditional attention mechanisms focus on aggregating information from multiple sources in which the modalities are aligned. However, they don't focus on the contrary information present in the modalities. We propose an approach that recognizes both harmonious and contrary information present in the different modalities. The contrary information present between the modalities should be identified for several reasons. Contrary pairs highlight differences and inconsistencies between modalities. They contribute to enhanced model interpretability and provide insights into why a model makes specific decisions. They are crucial for tackling problems with noisy or misaligned multimodal data, ensuring a stronger and more dependable analysis.
In this context, we introduce an attention mechanism called Inverted Dual Embedded Attention (IDEA). It combines harmonious and contrary attention vectors to reveal the complex interactions between different types of data. This unique strategy not only enhances the interpretability of the model but also provides a deeper understanding of multimodal information. By considering both harmonious and contrary attention, IDEA adapts to varying information needs, facilitating richer insights, improved representational learning, and a versatile approach to multimodal data analysis. 
IDEA contains four modules: (a) Shared Embedding Module; (b) Harmonious Attention Module; (c) Contrary Attention Module; and (d) Attended Fusion Module.
\paragraph{Shared Embedding Module:}

In the shared embedding module. We take both the textual and visual features in the same embedding space, say $d_e$=1024. Projecting textual and visual modalities into a shared embedding space offers several key advantages in a multimodal setting, such as seamless integration of information from text and images, effective cross-modal attention, enabling semantic alignment, and parameter sharing, which reduces model complexity, leading to more efficient training and improved generalization. 
Two types of modality transformation occur here: textual modality transformation and visual modality transformation.

\subparagraph{Textual Modality Transformation:}
In the initial stage, textual features $H_t \in \mathbb{R}^{d\times dt}$ undergo a non-linear projection into a shared embedding space. This involves a linear transformation using trainable parameters $W_t \in \mathbb{R}^{d_t \times d_{se}}$ and $b_t \in \mathbb{R}^{1 \times d_{se}}$, resulting in $H_t W_t + b_t$ as the linearly projected textual features within the shared embedding space.
The linearly projected textual features are subjected to a non-linear transformation denoted as $H_{{Text}} \in \mathbb{R}^{d\times d_{se}}$ that includes batch normalization~\citep{BN1,BN2} and a hyperbolic tangent activation.  Batch normalization helps in stabilizing and accelerating the training of deep neural networks by normalizing the inputs to each layer, reducing internal covariate shifts, and accelerating the training of deep neural networks. It normalizes the input of a layer by adjusting and scaling the activations. This can be shown in \cref{eqn:3}:

\begin{equation}\label{eqn:3}
{H_\text{Text}} = \tanh\left(\gamma \odot \left(\frac{(H_t W_t + b_t) - \mu}{\sqrt{\sigma^2 + \epsilon}} \right) + \beta\right)
\end{equation}

where $\gamma$ and $\beta$ are learnable parameters responsible for scaling and shifting, and $\mu$ and $\sigma$ represent the mean and standard deviation calculated over the mini-batch of the input data, respectively. $\varepsilon$ is a small constant added to the denominator in mathematical expressions to prevent division by zero or near-zero values, ensuring numerical stability. However, most of the deep learning frameworks, such as TensorFlow or PyTorch, have built-in functions for batch normalization. During training, these frameworks automatically calculate and update the mean and standard deviation over the mini-batches. 

\subparagraph{Visual Modality Transformation:}
This process is similar to that of textual modality. The only difference here is the transformation of the visual modality rather than the textual modality. In the initial stage, visual features $H_v \in \mathbb R^{d \times d_v}$ undergo a non-linear projection into a shared embedding space. This involves a linear transformation using trainable parameters $W_v \in \mathbb{R}^{d_v \times d_{se}}$ and $b_v \in \mathbb{R}^{1 \times d_{se}}$, resulting in $H_v W_v + b_v$ as the linearly projected visual features within the shared embedding space.

The linearly projected visual features are subjected to a non-linear transformation denoted as $H_{{Vis}} \in \mathbb{R}^{d\times d_{se}}$ that includes batch normalization \citep{BN1, BN2} and a hyperbolic tangent activation as expressed by the~\cref{eqn:4}:

\begin{equation}\label{eqn:4}
{H_\text{Vis}} = \tanh\left(\gamma \odot \left(\frac{(H_v W_v + b_v) - \mu}{\sqrt{\sigma^2 + \epsilon}} \right) + \beta\right)
\end{equation}

where $\gamma$, $\beta$, $\mu$ and $\sigma$ are already defined in \cref{eqn:3}
The dimensions of various parameters in our case are $d=128$, $d_{se}=1024$, $d_t=768$, and $d_v=1024$.
\subparagraph{Attention Matrix Computation:}

The attention matrix  $S_{{sim}}[i,j] \in \mathbb{R}^{d \times d}$ is a crucial intermediary representation that quantifies the degree of interaction between elements in two sequences, such as text and image embeddings. The computation of the attention score matrix involves taking the dot product of the text embeddings matrix and the transpose of the image embeddings matrix, followed by scaling the result by the square root of the dimensionality of the shared embeddings (\(d_{\text{se}}\)) as shown in the \cref{eqn:152}.
\begin{equation}\label{eqn:152} 
    S_{{sim}}[i,j] = \frac{{H_\text{Text}} \cdot {H_\text{Vis}}^T}{\sqrt{d_{\text{se}}}} 
\end{equation}

\begin{figure*}
    \centering
    \includegraphics[width=\linewidth]{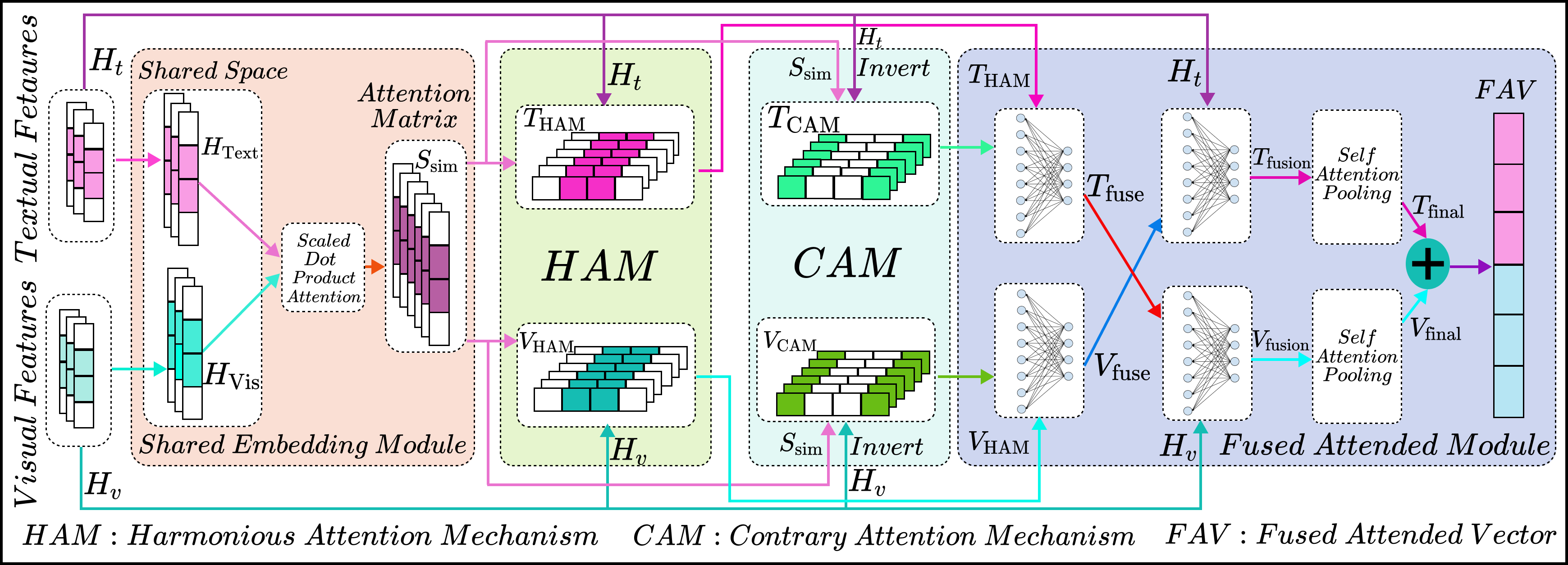}
    \caption{Inverted Dual Embedded Attention}
    \label{Fig.2}
\end{figure*}

The dot product operation captures the pairwise interactions between corresponding elements in the text and image sequences, while the scaling factor helps stabilize the softmax function during subsequent steps. 
The resulting attention matrix $S_{\text{sim}}[i, j] \in \mathbb{R}^{d \times d}$, every element in row \(i\) captures the degree of similarity between the \(i\)-th visual feature and the entire set of textual features. Correspondingly, each element in column \(j\) quantifies the similarity between the \(j\)-th textual feature and the entire set of visual features. This matrix serves as a foundation for generating attention weights and ultimately producing an attention-weighted sum of the original embeddings, facilitating the model's focus on relevant information in both sequences. 
Recent research studies \citep{hierarchical, attention} have predominantly focused on using attention matrices to evaluate co-attention weights by comparing similarities among different modalities. In contrast, our approach, ``Inverted Dual Embedded Attention (IDEA)'', does things differently. It calculates attention in a way that highlights differences and similarities, helping us understand how individual aspects of one type of information relate to the whole set of aspects of another type. It captures the hidden and complex information via two attention mechanisms, the Harmonious Attention Mechanism (HAM) and the Contrary Attention Mechanism (CAM), as shown in~\autoref{Fig.2}.

\paragraph{Harmonious Attention Mechanism (HAM):}
Harmonious attention mechanism emphasizes the alignment and correlation between features or information from different modalities. It is used to identify and highlight shared or similar aspects of the data. In a multimodal sentiment analysis task, harmonious attention is defined as the part that looks at words in a text description and elements in an accompanying image that expresses the same emotional sentiment. This aligns textual and visual cues associated with sentiment effectively. The harmonious attention vector facilitates the identification of highly correlated features between the two modalities. So, in a harmonious attention mechanism, we obtain the visual-guided textual features and vice versa. The visual-enriched textual features are obtained by applying the softmax function to the attention matrix. This process yields attention weights, which are subsequently multiplied by the textual features.
We introduce a temperature parameter $T$ in the softmax function to control the softness or sharpness of the probability distribution\citep{Temp}. A higher temperature encourages a softer, more uniform probability distribution, while a lower temperature value results in a sharper, more peaked probability distribution \citep{Temp2, Temp3}. For HAM, a higher temperature might lead to more evenly spread attention over the elements of the attention matrix, whereas a lower temperature might result in more focused attention on a few specific elements. After rigorous experimentation, the value of T for HAM is 1.65, and for CAM, it is 0.75. We obtain the attention weights or scores as expressed in \cref{eqn:154}:
   \begin{equation}\label{eqn:154}
   (Att_\text{weights})_{i, j} = \frac{\exp\left(\frac{S_{\text{sim}, i, j}}{T}\right)}{\sum_{k=1}^{d} \exp\left(\frac{S_{\text{sim}, i, k}}{T}\right)}
\end{equation}
$Att_\text{weights}$  is a matrix where each element $(Att_\text{weights})_{i, j}$ represents the attention weight for the corresponding element in the attention matrix $S_{\text{sim}}$ along axis 
$i$.
The visual-enriched textual features $T_\text{HAM}  \in \mathbb{R}^{d \times dt}$ can be computed as shown in the \cref{eqn:155}
\begin{equation} \label{eqn:155}
   T_\text{HAM} = \sum_{j=1}^d (Att_\text{weights})_{i, j} \cdot (H_t)_{j, k}
\end{equation}
$T_\text{HAM}$ is the attended representation matrix obtained after the attention mechanism. It represents the weighted combination of textual features $H_t$ based on the attention weights $(Att_\text{weights})_{i, j}$.

Similarly, the textual-enriched visual features  $V_\text{HAM} \in \mathbb{R}^{d \times d_v}$ are computed by employing the softmax function on the attention matrix to produce attention weights, which are then multiplied by the visual features. Since the attention weights $(Att_\text{weights})_{i, j}$ have already been computed in \cref{eqn:154}, we will use the same attention weights for the computation of $V_\text{HAM}$ as shown in \cref{eqn:156}. 

\begin{equation}\label{eqn:156}
{V_\text{HAM}} = \sum_{j=1}^d {(Att_\text{weights})_{i, j}} \cdot (H_v)_{j,k}
\end{equation}
In summary, the primary purpose of HAM is to capture and magnify the areas where information from different modalities aligns or agrees.
\paragraph{Contrary Attention Mechanism (CAM):}
Contrary attention highlights differences and disparities between features or information from different modalities. It is used to identify and accentuate areas where modalities diverge or provide contrary information. In a multimodal content moderation task, contrary attention might identify text and image content that contradict each other, helping the model detect potentially divergent content where the text and visual elements are in conflict. However, in most prior studies, contrary information, which can also be valuable for representation, is often overlooked. As shown in~\autoref{Fig.2}, we compute CAM using the attention matrix and the text and image modalities after multiplying them by -1 (designated as Invert).
Multiplying embeddings with -1 will reverse the direction of the embedding in the vector space~\citep{negative}. It does not alter the magnitude of the features. This can be useful for tasks such as learning to distinguish between real and fake data, generating adversarial examples, and finding various different hidden patterns.
The inverted visual-enriched textual features $T_\text{CAM} \in \mathbb{R}^{d \times dt}$ are computed as shown in \cref{eq:8}.
\begin{equation} \label{eq:8}
T_\text{CAM}  = \sum_{j=1}^d (Att_\text{weights})_{i, j} \cdot (-H_{t})_{j,k}
\end{equation}

Similarly, the inverted textual-enriched visual features  $V_\text{CAM} \in \mathbb{R}^{d \times dv}$ are computed as shown in \cref{eq:9}: 
\begin{equation} \label{eq:9}
V_\text{CAM} = \sum_{j=1}^d (Att_\text{weights})_{i, j} \cdot (-H_v)_{j,k}
\end{equation}
In the context of the contrary attention mechanism, it becomes imperative to use normalization of the feature vectors because of negation. However, the harmonious attention mechanism, being a standard attention mechanism without feature negation, may not require the same normalization considerations. In summary, the primary purpose of CAM is to capture the discrepancies, variations, or inconsistencies between modalities. It helps the model understand where and how modalities differ, contributing to a more nuanced analysis.
\paragraph{Fused Attended Module:}
To combine both harmonious and contrary attention vectors, we utilize dense layers akin to the structure found in multi-head attention~\citep{cross}, as illustrated below:
The textually-informed visual features are computed as shown in~\cref{eq:10}:
\begin{equation} \label{eq:10}
V_{fuse} = tanh(( V_\text{HAM} \oplus V_\text{CAM})W_\text{vis} + {b_\text{vis}})
\end{equation}
where $\oplus$ means concatenation operator, $V_\text{fuse} \in \mathbb{R}^{d \times dv}$, \(W_\text{vis} \in \mathbb{R}^{2dv \times dv}\) and $b_\text{vis} \in\mathbb{R}^{1 \times dv}$  are trainable parameters. \\
\begin{equation} \label{eq:11}
T_{fuse} = tanh(( T_\text{HAM} \oplus T_\text{CAM})W_\text{text} + {b_\text{text}})
\end{equation}
where $T_\text{fuse} \in \mathbb{R}^{d \times dt}$, \(W_\text{text} \in \mathbb{R}^{2dt \times dt}\) and  $b_\text{text} \in\mathbb{R}^{1 \times dt}$ are trainable parameters. After obtaining $T_\text{fuse}$ and $V_\text{fuse}$, We utilize dense layers to capture and understand the relationship that exists between the visual features and the enriched fused features. This process enables us to obtain what we refer to as the visual-fusion feature ($V_\text{fusion}$), as shown in \cref{vfusion}.
\begin{equation} \label{vfusion}
V_{fusion} = tanh((T_{fuse} \oplus H_{v})\widetilde{W_v} + \widetilde{b_v})
\end{equation}
$V_\text{fusion} \in \mathbb{R}^{d \times dv}$, $\widetilde{W_v} \in \mathbb{R}^{(dt+dv) \times dv}$, and ${\widetilde{b_v}} \in \mathbb{R}^{1 \times dv}$ are trainable parameters.

Similarly, to obtain the textual-fusion feature ($T_{fusion}$), we employ dense layers to capture the correlation between the textual features and the enhanced fused features, as shown in \cref{Tfusion}.
\begin{equation} \label{Tfusion}
T_{fusion}= tanh((V_{fuse} \oplus H_{t})\widetilde{W_t} + \widetilde{b_t})
\end{equation}
 $T_\text{fusion} \in \mathbb{R}^{d \times dt}$, $\widetilde{W_t} \in \mathbb{R}^{(dt+dv) \times dt}$ and  $\widetilde{b_t} \in \mathbb{R}^{1 \times dt}$ are trainable parameters. \\
After obtaining $V_\text{fusion}$ and $T_\text{fusion}$, the next step involves extracting the most significant representation from the combined textual and visual features. We apply global average pooling to the first dimension of $T_{\text{fusion}} \in \mathbb{R}^{d \times dt}$ and $V_{\text{fusion}} \in \mathbb{R}^{d \times dv}$, the dimensions reduce to $T_{\text{fusion}} \in \mathbb{R}^{1 \times dt}$ and $V_{\text{fusion}} \in \mathbb{R}^{1 \times dv}$ respectively. It is used to reduce dimensionality, prevent overfitting, and provide a summarized representation of the feature maps. We compute the attention scores on the pooled features using a self-attention mechanism~\citep{additive}.

The attention score ($e_i$) for $T_{\text{fusion}}$ is calculated using the scaled dot-product attention formula.
\begin{equation}\label{eqn:score}
e_i = \frac{(T_{\text{fusion}} W_Q)(T_{\text{fusion}} W_K)^T}{\sqrt{\text{dz}}} 
\end{equation}
In~\cref{eqn:score}, $W_Q \in \mathbb{R}^{dt \times dz }$ and $W_k \in  \mathbb{R}^{dt \times dz }$ are parameter matrices. 
The term \(\sqrt{\text{dz}}\) represents the scaling factor, where \(\text{dz}\) is the dimension of the query and key spaces.

The attention weight (\(\alpha_i\)) is obtained by applying the softmax function to the attention scores as shown in ~\cref{eqn:weights}:
\begin{equation}\label{eqn:weights}
\alpha_i = \frac{\exp(e_i)}{\sum_{k=1}^{n} \exp(e_k)}
\end{equation}
 where n denotes the number of elements in the input sequence.

The final textual fused feature vector representation $T_\text{final} \in \mathbb R^{1 \times dt}$ is a weighted sum of the values based on attention weights as shown in \cref{final}:
\begin{equation}\label{final}
T_\text{final}= \sum_{k=1}^{n} \alpha_k (T_{\text{fusion}} W_V) 
\end{equation}
where $W_V \in \mathbb{R}^{\text{dt} \times \text{dz}}$ is parameter matrix and \text{dz} 
is the output dimensionality, which matches that of $dt$.

Similarly, for $V_\text{fusion}$, repeat~\cref{eqn:score,eqn:weights,final}, by replacing $T_\text{fusion}$ with $V_\text{fusion}$, we can get the final representation of visual fused feature vector as $V_\text{final} \in \mathbb R^{1 \times dv}$.

After obtaining the final textual-fused feature vector $T_{\text{final}}$ through the attention mechanism, the last step involves concatenating it with the final visual-fused feature vector $V_{\text{final}}$. This results in a single fused attended vector (FAV)$ \in \mathbb{R}^{1 \times (dt+dv)}$ that captures the most significant representations from both textual and visual features.
The concatenation operation is shown in \cref{concatenation}.
\begin{equation}\label{concatenation}
    FAV= {T}_\text{final} \oplus {V}_\text{final}
\end{equation}
The fused attended vector combines the enriched information from both modalities, providing a comprehensive representation for downstream tasks such as classification. This integrated feature captures the intricate relationships between textual and visual features, enhancing the model's understanding and performance.

In summary, IDEA takes into account the similarities and dissimilarities of the textual and visual features, which in turn captures the rich representation of textual-guided visual features and visual-guided textual features. Its unique ability to discern both similar and dissimilar features enables a more comprehensive understanding of multimodal data. By dynamically allocating attention to relevant aspects in text and image inputs, IDEA excels in capturing subtle relationships, allowing the model to operate with a heightened level of context awareness. This dual capability of handling similarity and dissimilarity in patterns positions IDEA as a versatile tool in tasks such as multimodal fusion, where nuanced connections between different modalities play a crucial role. The model's adaptability to diverse patterns fosters a richer representation, enhancing its performance across a spectrum of applications. IDEA's contribution underscores the importance of attention mechanisms that go beyond the traditional focus on similarities alone. By accommodating both ends of the spectrum, IDEA serves as a testament to the evolving landscape of attention mechanisms, pushing the boundaries of what is achievable in capturing the complexities of multimodal information.

\subsection{Multimodal Graph Learning} 
We delve into the utilization of graph-based algorithms for learning representations of both text and image data. Multimodal Graph Learning is a technique that combines graph-based representations with multiple data modalities, such as text and images. It involves constructing a graph where nodes represent different types of data and edges capture relationships. The two sub-modules of the Multimodal Graph Learning module are Multimodal Encoder and Graph Knowledge Propagation.

\subsubsection{Multimodal Encoder}
We discuss the feature extraction process of a multimodal encoder to create shared embedding space for both images and text. The multimodal encoder comprises two primary components: a text encoder and an image encoder. These components generate embeddings that capture the semantic relationships between images and corresponding textual descriptions. The image encoder is responsible for transforming input images into contextually enriched embeddings. On the other hand, the text encoder processes textual descriptions to produce corresponding embeddings. Utilizing transformer-based architectures, it captures contextual relationships within the text. The primary aim of the multimodal encoder is to align the embeddings of images and text within a shared semantic space. This alignment is achieved through a contrastive learning objective that encourages similar image and text pairs to have closer embeddings while pushing dissimilar pairs apart. We use a multimodal CLIP encoder \citep{CLIP} for this task.

\subsubsection{Graph Knowledge Propagation}
Graph-based knowledge propagation is the technique of disseminating information through a graph structure, where nodes signify features or data points, and edges depict relationships among them. It plays a pivotal role in tasks like information dissemination and graph-based learning. We present Algorithm \ref{alg:GCAKP}, an approach for the construction of graphs and their knowledge propagation. The algorithm facilitates an iterative refinement of node representations of both image and text graph features. It encompasses a detailed procedure that includes the construction of graphs and the subsequent distribution of knowledge across various layers of the graph. In Algorithm \ref{alg:GCAKP}, the operations from lines 1 to 19 are centered on constructing a similarity matrix ($A$). This process entails evaluating the cosine similarity between pairs of feature vectors denoted as $H_i$ and $H_j$. The decision to establish an edge between feature vectors depends on whether the computed cosine similarity $S_{ij}$ surpasses a predefined threshold (threshold $\geq$ 0.75). The resulting undirected edges are then incorporated into the adjacency matrix $A$. Consequently, lines 31 and 32 lead to the creation of two adjacency matrices: $A_i$, known as the Image similarity matrix, and $A_t$, known as the text similarity matrix. Subsequently, from lines 20 to 30, we employ GraphSAGE (Graph Sample and Aggregated) \citep{GraphSage} for the propagation of knowledge across the constructed matrices. It is a node embedding algorithm designed specifically to acquire representations for nodes within a graph by aggregating information from their neighboring nodes. It strikes a balance between capturing local structural information and leveraging feature information associated with nodes. The fundamental concept behind this is to generate node embeddings by gathering information from the neighboring nodes of a given node. Here, each node represents a text feature in a text similarity matrix or an image in an image similarity matrix.
\begin{algorithm}[!ht]
\caption{Graph Construction and Knowledge Propagation} \label{alg:GCAKP}
\textbf{Input:}
Image Feature $H_i$ (size $n$),Text Feature $H_t$ (size $n$), n( $V_i$)= n($V_t$),
$K$: Number of layers or depth,threshold=0.75, neighborhood function $N : V \to 2^v$.\\
\textbf{Output:}
Updated node embeddings for text and image features $ \forall v \in V_i$ and $\forall v \in V_t$
\begin{algorithmic}[1]
\Function{SimilarityGraph}{${H_i}$,threshold}
    \State Initialize adjacency matrix $A$ with zeros
    \For{$i = 1$ to $n$}
        \For{$j = i + 1$ to $n$}
            \State Let $H_i$ be the $i$-th feature vector 
            \State Let $H_j$ be the $j$-th feature vector
            \Statex Calculate cosine similarity between $H_i$ and $H_j$:
            \State $S_{ij} = \text{cosine\_similarity}(H_i, H_j)$

            \If{$S_{ij} > \text{threshold}$}
                \State Add an edge between $i$-th and $j$-th feature vector in the adjacency matrix:
                \State $A[i][j] = 1$
                \Statex As the graph is undirected, set the symmetric entry:
                \State $A[j][i] = 1$
            \Else
                \Statex No edge between  $i$-th and $j$-th feature vector below threshold:
                \State $A[i][j] = 0$
                \State $A[j][i] = 0$
            \EndIf
        \EndFor
    \EndFor
    \State \textbf{Return} $A$
\EndFunction
\Function{KnowledgeGraphPropagation}{$A,V, K$}
    \State Initialize node embeddings $h_v^0 \leftarrow (A)_v, \forall v \in V$
    \For{$k = 1$ to $K$}
        \For{$v \in V$}
            \Statex Sample $N(v)$ from the neighborhood of node $v$ using the adjacency matrix $A$
            \State $Agg^k_{N(v)} \leftarrow \text{Aggregate}_k({h_{u}^{k-1}, \forall u \in N(v)})$
            \State $h_v^k \leftarrow \sigma\left(W_k \cdot \text{Concat}(h_v^{k-1}, Agg^k_{N(v)})\right)$
        \EndFor
        \State Normalize embeddings: $h_v^k \leftarrow \frac{h_v^k}{\lVert h_v^k \rVert_2}, \forall v \in V$
    \EndFor
    \State \textbf{Return} Updated node embeddings $h_v^K$ for all $v \in V$
\EndFunction

\State $A_i \gets$ \textsc{SimilarityGraph}($H_i$, threshold)
\State $A_t \gets$ \textsc{SimilarityGraph}($H_t$, threshold)
\State $I_v \gets$ \textsc{KnowledgeGraphPropagation}($A_i$, $V_i$, $K$)
\State $T_t \gets$ \textsc{KnowledgeGraphPropagation}($A_t$, $V_t$, $K$)

\end{algorithmic}
\end{algorithm}

For a given node \(v\) (representing a text or an image feature) and a specific layer \(k\), the embedding generation process in GraphSAGE consists of the following steps, as depicted in lines 21-30: Neighbor Sampling, Aggregation, and Encoding. The process commences with the initialization of node embeddings $h_0^v$, instantiated from the similarity matrix ($A$). Subsequently, the algorithm systematically advances through $K$ layers, iteratively refining node representations. For each node $v$, a set of neighbors $N(v)$ is sampled from the adjacency matrix $A$. Information from the local neighborhood is then aggregated using a designated function ${Aggregate_k}$.
The aggregation function can be as simple as using the mean to capture neighborhood information. This can be shown in~\cref{eq:12}.
\begin{equation}\label{eq:12}
 Agg^k_{N(v)}\leftarrow \text{Aggregate}_k(\{h_{u}^{k-1}, \forall u \in N(v)\})
\end{equation}
where, \(h_u^{k-1}\) is the embedding of neighbor \(u\) at layer \(k-1\), and \(\text{N}(v)\) is the set of sampled neighbors of node \(v\).
The updated node embeddings $h_v^k$ are computed by concatenating the aggregation result with the prior embeddings, facilitated by a trainable weight matrix $W_k$. 
 The purpose of this step is to update the embedding of the central node by considering both its local neighborhood information and its own features. This can be represented as shown in~\cref{eq:13}: 
\begin{equation} \label{eq:13}
h_v^k \leftarrow \sigma\left(W_k \cdot \text{Concat}(h_v^{k-1}, Agg^k_{N(v)})\right)
\end{equation}
where, \(\sigma\) is an activation function (e.g., ReLU), \(W_k\) is the weight matrix for the \(k\)-th layer. This process is repeated for multiple layers \(k\) to capture different scales of neighborhood information. This process is repeated for every node till we get the updated feature of each node. The updated node embeddings are normalized after each layer to prevent gradient explosion and enforce stability and convergence. This iterative refinement process is reiterated over $K$ layers, resulting in progressively enhanced node representations.

The ultimate output of this process manifests in the form of updated node embeddings for both image and text documents or nodes. The updated image node embeddings are denoted as $I_v$, as can be seen in line 33, while the corresponding updated text node embeddings are represented as $T_v$ as can be seen in line 34. These embeddings capture nuanced representations that encapsulate information extracted from the constructed image and text similarity graphs.
In summary, GraphSAGE serves as a method to aggregate embeddings from neighboring nodes for a specific target node. Each iteration results in the generation of updated node representations for every node in the graph. The application of multiple stacked layers facilitates the creation of intricate structural and semantic-level features, contributing to enhanced capabilities for various downstream tasks.

\subsubsection{Graph Fusion Learning Network (GFLN)}
In the Graph Fusion Learning Network, the fusion of updated text and image graph features into a unified representation is achieved through concatenation. This is done to preserve discriminative information about each modality. The choice for this concatenation approach is rooted in the fact that we aim to obtain two types of features, i.e., cross-modal features and modality-specific refined features. Cross-modal features are obtained through the IDEA mechanism, enabling the model to capture and highlight inter-modal relationships. Concurrently, the graph learning module refines modality-specific features, emphasizing unique characteristics within each modality. Due to this strategy, the model ensures the preservation of both cross-modal interactions and modality-specific nuances, contributing to a comprehensive and enriched representation of the integrated information. Following the concatenation process, the unified vector achieves a dimensionality of 1024. Subsequently, we utilize two consecutive fully connected layers: the first layer reduces the dimensionality to 512 with a dropout rate of 0.2, followed by a second layer, further reducing it to 256 with the same dropout rate. This processed vector is subsequently fed into the multimodal fusion module, enabling the seamless integration of information from diverse sources.

In summary, the process begins by constructing two matrices or graphs based on cosine similarity, one for each modality. These graphs serve as input to the GraphSAGE algorithm, which updates node embeddings through iterative operations. The updated node embeddings from both modalities are fused using concatenation, preserving modality-specific features. The resulting unified vector, capturing enriched representations from both graphs, is then forwarded to the multimodal fusion module for further integration of information from distinct modalities.

\subsection{Social Context Feature Extraction} \label{SCF}
In various existing works, the predominant focus has been placed solely on textual and visual features. To the best of our knowledge, none of them has extensively used Social Context Features in their model in multimodal settings to enhance crisis response. Our approach involves investigating the collective influence of textual, visual, and Social Context Features for identifying crisis-related information. Social Context Features (SCF) encompass a range of attributes, including crisis-related posts, user interaction history, etc. We have divided Social context Features into two parts: Text Semantic Profiling (TSP) and Social Interaction metrics (SIM).

\subsubsection{Text Semantic Profiling}
Text Semantic Profiling (TSP) is a technique that aims to decipher the contextual and conceptual facets embedded within text.TSP helps us to extract and categorize semantic attributes such as sentiment, emotions, topics, relationships, and intent.TSP includes the EmoQuotient Analysis, SentiQuotient Analysis, and Crisis Informative Score.

\paragraph{SentiQuotient Analysis (SentiQ):} 
SentiQ helps in deciphering the text's subjective tone, thus contributing to a holistic interpretation of its underlying connotations. We use VADER (Valence Aware Dictionary and Sentiment Reasoner) \citep{VADER} to extract the sentiments of the text. It assigns a sentiment score to each word in a piece of text. It computes a compound score that reflects the overall sentiment of the text, spanning from highly negative to highly positive, with neutrality positioned in the middle. It also provides scores for positive, negative, and neutral sentiments. By incorporating sentiment scores, models can better differentiate between classes by considering contextual sentiment and potentially weighing emotionally charged words more effectively. The SentiQ is represented as follows:
\begin{equation}\label{eq:16}
SentiQ = Sentiment\_Encoder(T_i)
\end{equation}
where, $T_i$ denotes the tweets for which we must determine the sentiment expressed in the text.

\paragraph{EmoQuotient Analysis (EmoQ):} 
EmoQuotient Analysis focuses on discerning emotional nuances expressed within the text, like happiness, sadness, anger, fear, etc. We use EmoLex \citep{Emolex} to quantitatively assess emotional content within textual data. Emotion Quotient plays a significant role in text classification by providing additional context and information about the emotional tone of the text. It gives 13 labels for a particular text \citep{Anshul}. Due to this, we will get fine-grained scores for a text. These scores enhance the accuracy and granularity of the classification task. The EmoQ is represented as:
\begin{equation}\label{eq:17}
EmoQ = {Emotion\_Encoder}(T_i)
\end{equation}
where, $T_i$ signifies the tweets for which we need to grasp the emotional tone conveyed in the text.

\paragraph{Crisis Informative Score (CIS):}
Social media text often contains spelling errors, abbreviations, user mentions, and numerous hashtags. It becomes crucial to employ a metric for categorizing informative or humanitarian content within these posts. This metric, Crisis Informative Score (CIS), assesses the semantic characteristics of text, offering a quantitative measure of crisis-related terminology within the text. To compute CIS as outlined in Algorithm~\ref{alg:User_Crisis_Informative_Score}, in line 8, we initiate the crisis lexicon count at zero. For each tweet in the dataset, we compare its content to a carefully curated lexicon containing 4,268 crisis-related terms. The number of words in a single tweet that match the crisis lexicon is tallied, as can be depicted from lines 9-15. After this process across all tweets in the dataset, we perform min-max normalization on the count of matched words within the tweets, as demonstrated in lines 16-19 of the CIS function in Algorithm~\ref{alg:User_Crisis_Informative_Score}. The CIS is represented as follows:
\begin{equation}\label{eq:18}
CIS= normalise(Crisis\_Lexicon\_Count)
\end{equation}

where, \textit{normalise} denotes the min-max normalization of the crisis-lexicons count present in the tweet.

\subsubsection{Social Interaction Metrics (SIM)} 

Social Interaction Metrics are essential for understanding and analyzing how users engage with content and each other. Two significant aspects of these metrics are User Engagement Metrics (aka User Engagement Features) and User Informative Score.

\paragraph{User Engagement Metrics (UEM):} 
User Engagement Features encapsulate the ways in which individuals interact with content on social media. These features provide insights into the effectiveness of user content. Common user engagement features include Likes, Retweets, Replies, Followers, Friends,  etc. We normalize all these scores

\begin{algorithm}
\caption{User-Crisis Informative Score}
\label{alg:User_Crisis_Informative_Score}
\textbf{Input:}
\\
$U_i$: Number of informative tweets posted by a user $U$ \\
$U_{ni}$: Number of non-informative tweets posted by a user $U$ \\
$\mathcal{U}_{\text{Total}}$: Total number of tweets posted by a user $U$ 

\textbf{Output:}
\\
$\phi_c$: Crisis Informative Score (CIS) \\
$\phi_u$: User Informative Score (UIS) \\
$\phi_{u_c}$: User Crisis Informative Score (UCIS)

\begin{algorithmic}[1]
\Function{UIS} {$\mathcal{U}_{\text{Total}}$, $U_i$, $U_{ni}$}
    \State $\phi_u \gets \text{user\_informative\_score}(\mathcal{U}_{\text{Total}}, U_i, U_{ni})$
    \State $\phi_u \gets \frac{U_i - U_{ni}}{\mathcal{U}_{\text{Total}}}$
    \State $\widetilde{\phi}_u  \gets \frac{\phi_u - \min(\phi_u)}{\max(\phi_u) - \min(\phi_u)}$
    \State \textbf{return} $\widetilde{\phi}_u$
\EndFunction

\Function{CIS} {$\widetilde{T_i}$, $\mathcal{L}_{\text{lex}}$} // $\widetilde{T_i}$ represents all the tweets present in the dataset.
    \State $count\_crisis \gets 0$ 
    \For{$\text{t}$ \textbf{in} $\widetilde{T_i}$} // t is single tweet
        \For{$\text{w}$ \textbf{in} $\text{t}$}    // w represents single word in tweet
            \If{$\text{w}$ \textbf{is in} $\mathcal{L}_{\text{lex}}$} // $\mathcal{L}_{\text{lex}}$ is lexicon list.
                \State $count\_crisis \gets count\_crisis + 1$ 
            \EndIf
        \EndFor
    \EndFor
    \State $\phi_c \gets \text count\_crisis$
    \State $\widetilde{\phi}_c  \gets \frac{\phi_c - \min(\phi_c)}{\max(\phi_c) - \min(\phi_c)}$
    \State \textbf{return} $\widetilde{\phi}_c$
\EndFunction

\Function{UCIS} {$\phi_{u_c},\widetilde{\phi}_c,\widetilde{\phi}_u$}
    \State $\phi_{u_c} \gets \alpha \cdot \widetilde{\phi}_u + (1 - \alpha) \cdot  \widetilde{\phi}_c$
    \State \textbf{return} $\phi_{u_c}$
\EndFunction
\end{algorithmic}
\end{algorithm}

\paragraph{User Informative Score (UIS):} 
The User Informative Score is a metric designed to assess the credibility and informativeness of a user's content. It is determined by considering the following factors: the total number of tweets posted by a user denoted as $U_\text{Total}$, the number of informative tweets posted by a user denoted as $U_\text{Info}$, and the number of non-informative tweets posted by a user, represented as $U_\text{Non-Info}$. As depicted in Algorithm~\ref{alg:User_Crisis_Informative_Score}, lines 1-5 tell about the computation of the UIS. It can be represented as shown in~\cref{eq:19}.

\begin{equation}\label{eq:19}
UIS = \frac{U_{{Info}} - U_{{Non-Info}}}{U_{{Total}}}
\end{equation}
The UIS metric is then normalized using min-max normalization.
The Social Holistic Vector (SHV) comprises all the scores we got from \cref{eq:16}, ~\cref{eq:17}, ~\cref{eq:18}, and ~\cref{eq:19}. 

\begin{equation}\label{eq:20}
SHV = {Concat}(\psi_e,\psi_s ,\psi_u, \psi_ui, \psi_ci)
\end{equation}
As shown in~\cref{eq:20}, $\psi_e, \psi_s, \psi_u, \psi_ui, \psi_ci$ represent the Emotion Quotient, Sentiment Quotient, User Engagement Metrics, User Informative Score, and Crisis Informative Score respectively.
\paragraph{User-Crisis Informative Score (UCIS):} 
 It is important to highlight that we employ a weighted sum approach, outlined in Algorithm \ref{alg:User_Crisis_Informative_Score}, to combine the Crisis Information Score (CIS) and User Information Score (UIS). The weighted sum approach, where each score is assigned a specific weight, ensures a balanced consideration of the dual aspects, with CIS capturing the relevance of crisis content and UIS gauging the credibility of the user. This method provides flexibility by allowing adjustments to the weights, reflecting the relative importance of each aspect based on specific assessment goals or evolving contextual priorities. The resulting combined score offers a more nuanced and comprehensive measure, ultimately improving the model's ability to discern reliable and informative content during crises.

\subsection{Multimodal Fusion Network (MFN)}
In this section, we elaborate on the Multimodal Fusion Network.~\autoref{Fig.1} illustrates different constituents of the MFN module. This module comprises four sub-modules, namely, (a) Multimodal Adaptive Learning Network, (b) Social Contextual Learning Network, (c) Joint Fusion Learning Network, and (d) Prediction Layer.
\subsubsection{Multimodal Adaptive Learning Network (MALN)} In this module, the fused attended vector (FAV) further refines and distills the integrated features by learning higher-level patterns and relationships that are specific to the classification task. The network can extract informative features and establish inter-dependencies between textual and visual features, ultimately enhancing the predictive capabilities of the model. The Fused Attended Vector (FAV), initially with dimensions of 2048, undergoes a hierarchical reduction in dimensionality. It passes through successive fully connected layers, achieving dimensions of 1024, 512, and 256, respectively. Each layer incorporates a dropout rate of 0.2. The refined output is then directed to the Joint Fusion Learning Network for further processing.\\

\subsubsection{Social Holistic Learning Network (SHLN)} 
The Social Holistic Vector (SHV) consolidates semantic text information and social interaction information, where the latter encompasses user engagement information and user-generated informative content shared on social media platforms. The user engagement data, including favorites count, retweet count, followers count, friends count, and status count, undergoes initial scaling using min-max normalization. This processed vector is referred to as the user engagement metrics. Similarly, user-generated informative content undergoes scaling through min-max normalization, yielding the User Informative Score (UIS), as outlined in~\autoref{SCF}. The semantic textual information encompasses emotional, sentimental, and Crisis Informative Score (CIS). Following the concatenation of these processed feature vectors, the resulting comprehensive representation is denoted as the Social Holistic Vector, characterized by dimensions totaling 21. Specifically, it consists of 3 dimensions for sentiment, 11 for emotion, 1 for CIS, 5 for user engagement, and 1 for UIS. SHV encapsulates a holistic view of user interaction and content-related information on social media platforms. The SHV undergoes an initial processing stage by traversing through a fully connected layer, where its dimensions are effectively reduced to 16. To mitigate overfitting, a dropout of 0.2 is applied during this layer. These transformation includes dimensionality adjustments and the introduction of non-linearity through activation functions such as  ReLU, enabling the network to capture intricate data patterns. During this process, the network extracts high-level features relevant to the classification task, including patterns related to user engagement, informative content, emotional and sentimental aspects. The resulting output represents a refined and condensed representation of the combined social and text information. This output is given to the Joint Fusion Learning Network.\\

\subsubsection{Joint Fusion Learning Network (JFLN)} 
The feature vectors generated by MALN, SHLN, and GFLN are concatenated for joint representation. This integration enables the neural network to comprehensively capture the essence of multimodal data by merging diverse information streams. The joint representation encodes inter-modality relationships and interactions, allowing the model to discern complex patterns and correlations that may not be apparent when considering each modality in isolation. The joint representation of feature vectors passes through several fully connected layers; their dimensions undergo a step-wise reduction from 528 to 256, implementing a dropout of 0.2. This is succeeded by the application of another layer, further reducing the dimensions to 128 while maintaining a dropout rate of 0.2. The subsequent fully connected layer continues this reduction to 64 dimensions, maintaining the same dropout rate. This consolidated representation of the feature vectors is optimized for the classification task, empowering the model to make more accurate predictions by leveraging collective knowledge from all modalities, including the social context.\\

\subsubsection{Prediction Layer} In the process of classifying content for the Informative Task, the model generates predicted class labels by passing the JFLN module's output to the prediction layer. This layer employs a Sigmoid activation function to compute predicted probabilities, mapping real-valued inputs to probabilities between 0 and 1. Model performance is assessed using binary cross-entropy loss, and the AdamW optimizer updates model weights based on loss gradients. The predicted probabilities are subsequently transformed into binary labels (0 or 1) using a thresholding function. If the probability surpasses or equals the threshold $\geq$ 0.5, it's assigned the label 1, denoting the Informative class; otherwise, it's labeled 0, indicating the non-informative class. This binary classification approach effectively distinguishes between informative and non-informative content.\\
\begin{equation}\label{binarcross}
L = -\frac{1}{N} \sum_{i=1}^{N} \left(y_i \log(p_i) + (1 - y_i) \log(1 - p_i)\right)
\end{equation}
As shown in \cref{binarcross}, $N$ represents number of examples, your dataset, $y_i$ represents true binary label (0 or 1) for the $i$-th example, $p_i$ represents predicted probability that the $i$-th example belongs to class 1.\\
Similarly, for the Humanitarian Task, a multiclass classification challenge, we implement a Softmax activation function in the prediction layer to establish a probability distribution across multiple classes. The resulting probability vector assigns a probability value to each class. Model performance is evaluated using a suitable loss function for multiclass classification, such as categorical cross-entropy loss. 
\begin{equation}\label{categoricaLCROSS}
L = -\frac{1}{N} \sum_{j=1}^{N} \sum_{i=1}^{C} y_{ij} \log(p_{ij})
\end{equation}
As shown in \cref{categoricaLCROSS}, $C$ represents number of classes, $N$ represents Number of examples, $y_i$ represents binary indicator (0 or 1) of whether class $i$ is the correct classification for the example, $p_i$ represents predicted probability assigned to class $i$ by the model, $y_{ij}$ represents binary indicator for class $i$ in the $j$-th example, $p_{ij}$ represents predicted probability for class $i$ in the $j$-th example.

The Adam optimizer calculates gradients for loss-based weight updates and model training. In practice, the class with the highest probability in the predicted distribution is selected as the model's classification decision. This approach enables the model to effectively categorize data into one of the available classes.

In summary, CrisisSpot is delineated into four integral components for the identification of crisis-related information. The primary Feature Extraction and Interaction Module orchestrates the interaction between text and visual modalities, extracted from text and image encoder, respectively, utilizing an IDEA attention mechanism. This mechanism adeptly captures both harmonious and contrary patterns, forwarding the extracted features to the subsequent Multimodal Fusion Module. The second component is dedicated to Multimodal Graph Learning, where features extracted from a multimodal encoder undergo enrichment of node features via a GraphSAGE layer. The resultant fused vector is channeled into the Multimodal Fusion Module. The third module is centered on Social Context Feature Extraction, encompassing diverse attributes such as crisis-related posts and user interaction history. This module includes metrics such as User Informative Score (UIS), Crisis Informative Score (CIS), and User Engagement. UIS evaluates user credibility, CIS quantifies crisis-related vocabulary presence, and User Engagement metrics capture pertinent text and user-centric information. The amalgamated features from this module are concatenated and fed into the Multimodal Fusion Module. The final module, Multimodal Fusion, processes the fused attended vector from the Feature Extraction Module and is passed through a dedicated Deep Neural Network, while a separate neural network handles the social holistic vector from the Social Context Feature Extraction Module. Outputs from these networks, in addition to graph-attended information, are further integrated through a Joint Fusion Learning Network, ultimately culminating in the prediction layer. This intricate architectural framework maximizes the synergistic utilization of textual, visual, and Social Context Features, augmenting the model's prowess in discerning crisis-related information.

\section{Experimental Evaluations} \label{Section.4}
We present an overview of the datasets employed for experimental evaluations. Subsequently, we discuss unimodal and multimodal state-of-the-art methods for comparative analysis. Additionally, we discuss the standard evaluation metrics used to compare the performance of different approaches.
 
\subsection{Datasets}
We provide an overview of the CrisisMMD dataset. Subsequently, we introduce the data curation and annotation process for our newly created dataset, TSEqD (Turkey-Syria Earthquake Dataset).
\subsubsection{CrisisMMD}
The CrisisMMD dataset~\citep{CRISISMMD} is a valuable and comprehensive resource in the field of disaster management. It consists of a vast collection of multimodal social media posts that encompass texts and images from seven natural disaster events. The CrisisMMD dataset serves as a benchmark for evaluating the performance of state-of-the-art methods in disaster management. The number of samples from the CrisisMMD dataset utilized in our study is displayed in \autoref{Table1}.

\begin{table}[pos=ht]
  \caption{Data Distribution for Informative and Humanitarian Tasks in CrisisMMD dataset} \label{Table1}
  \centering
  \begin{tabular}{p{1.5in}lccc}
    \toprule
  \textbf{Task} & \textbf{Train} & \textbf{Validation} & \textbf{Test}\\
    \midrule
    Informative & 9601 & 1573 & 1534 \\
    Humanitarian  & 6126 & 998 & 955 \\
    \bottomrule
  \end{tabular}
\end{table}

\subsubsection{TSEqD} Due to the dearth of multimodal datasets in the field of disaster management. We have curated a vast collection of social media posts comprising of texts and images from the Turkey-Syria Earthquake\footnote{{https://github.com/Shahid-135/CrisisSpot}}. The dataset, TSEqD (Turkey-Syria Earthquake Dataset), has two different sets of annotations: one for Informative Tasks and another for Humanitarian Tasks. We have chosen to adhere to the CrisisMMD dataset's consistent labeling format. The process of data curation, annotation, and the distribution of labels across various classes have been elaborated in subsequent sections.

\paragraph {Data Curation:} We annotated a total of 10,352 tweets obtained during the period from February 6, 2023, to March 16, 2023. Our objective was to curate a dataset that captures diverse facets of the disaster, including details about affected individuals, losses resulting from destruction, and ongoing efforts to assist affected individuals. We utilized the Twitter API to collect these tweets, including their textual content, associated images, and metadata related to the tweets. This comprehensive dataset can serve as a valuable resource for future tasks that may leverage user-specific information or utilize tweet metadata for analysis and research purposes. The number of text and image samples per label in the TSEqD dataset is shown in \autoref{Table2}. The label distribution within the TSEqD dataset is illustrated in~\autoref{Table3}.

\begin{table}[pos=ht]
  \caption{Number of Text and Image Samples per Label in TSEqD Dataset} \label{Table2}
  \centering
  \begin{tabular}{p{1.75in}lccc}
    \toprule
    \textbf{Description} &\textbf{Label} &\textbf{Text} & \textbf{Image} \\
    \midrule
    Informative & 1 & 7136 & 5363 \\
    Non-Informative & 0  & 3199 & 4972 \\
    \bottomrule
  \end{tabular}
\end{table}  
\paragraph{Multitask Annotation:}
We will begin by examining the annotation process for the Informative task (Task 1), which involves two labels: informative and non-informative. Following that, we will address the Humanitarian task (Task 2), which encompasses eight labels as discussed in subsequent sections.

\subparagraph{Annotation Process for Task 1 (Informative Task):} This task typically involves annotating posts to determine whether they contain valuable information, news updates, or relevant details regarding the ongoing crisis event. Informative Task annotations are essential for distinguishing between posts that contribute to situational awareness and those that do not, aiding in the effective management of disaster response efforts. Examples of such posts are the appeals of organizations asking for support to help the victims, aid camp information, information about a lost or found individual, and reports of death tolls. The post which does not contain any disaster-related information is annotated as non-informative.\\

\begin{table}[pos=ht]
  \caption{Distribution of Labels in TSEqD Dataset} \label{Table3}
  \centering
  \begin{tabular}{p{2.5in}lcc}
    \toprule
    \textbf{Label} & \textbf{Count} \\
    \midrule
    Both Informative & 4312 \\
    At least one Informative & 8187 \\
    Non-Informative & 2148 \\
    \bottomrule
  \end{tabular}
\end{table}

\subparagraph{Annotation Process for Task 2 (Humanitarian Task):}
\noindent In this task, we assign labels to each tweet based on its corresponding category. We elaborate on the different categories as follows:\\
\text{Infrastructure and Utility Damage:} This category shows the damage to built structures, e.g., Damaged roads, destroyed buildings, rubble, and many more.\\
\text{Vehicle Damage:} This category shows damaged vehicles such as broken windows and cars under debris.\\
\text{Rescue or Donations Efforts:} This category shows donations or volunteering efforts, e.g., Transport people to safe places, evacuate, and provide medical aid.\\
\text{Injured or Dead People:} This category shows the information related to injured or dead people. These involve current death tolls, injuries, and deaths seen in an area describing their cause.\\
\text{Affected Individuals:} This category shows things such as people sitting outside, waiting in line to receive aid, needing shelter, etc.\\
Missing or Found People: This category shows instances of missing or found people.\\
Other Relevant Information: This category shows some relevant information not covered by the categories mentioned above. These may include maps of the affected areas and possible attempts of fraud or criminal activities taking place using the disaster as a cover. Also, any images that have contact information or QR codes for donations can also be included.
\begin{table}[pos=ht]
  \caption{Distribution of Text and Image Samples in Each Class in TSEqD Dataset} \label{Table4}
  \centering
  \begin{tabular}{p{1.75in}lcccc}
    \toprule
   \textbf{Description} & \textbf{Label} & \textbf{Text} & \textbf{Image} \\
    \midrule
    Infrastructure and Utility damage & 1 & 131 & 1279 \\
    Vehicle Damage & 2 & 0 & 46 \\
    Rescue and donation & 3 & 4235 & 1659 \\
    Injured or dead people & 4 & 1182 & 199 \\
    Affected Individuals & 5 & 492 & 1125 \\
    Missing or found people & 6  & 83 & 12 \\
    Other relevant info & 7 & 1015 & 1041 \\
    Not relevant or can't judge & 8 & 3197 & 4974 \\
    \bottomrule
  \end{tabular}
\end{table} 
The distribution of text and image samples in each class in the TSEqD dataset is shown in~\autoref{Table4}
In brief, we have annotated the dataset for various tasks, such as the Informative Task (Task 1) and Humanitarian Task (Task 2). The annotated dataset, namely TSEqD, contains about 10,352 data samples. The dataset contains 22 columns and contains necessary meta information that can be helpful to the researchers. To the best of our knowledge, no such dataset in disaster management has this unique and meaningful data.

\begin{figure}[pos=ht]
    \centering
    \includegraphics[width=\linewidth]{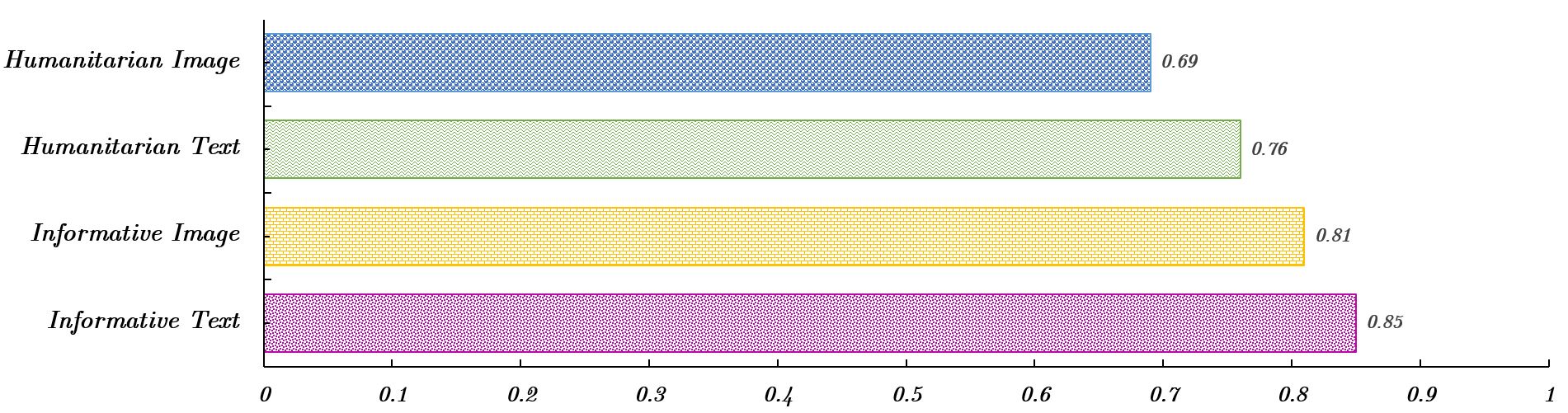}
    \caption{Inter-Annotator Scores for Informative and Humanitarian Tasks}
    \label{Fig.3}
\end{figure}

In summary, we have performed annotations for two distinct tasks: the Informative Task and the Humanitarian Task. The TSEqD dataset comprises approximately 10,352 data samples and possesses 22 columns, including essential metadata, which can be of great assistance to researchers. This dataset represents a valuable resource in the field of disaster management, offering unique and meaningful data.
\paragraph{Inter-Annotator Agreement:}
In the annotation process, we conducted separate annotations for both image and text data.
The inter-annotator agreement scores for different modalities under different tasks can be seen in~\autoref{Fig.3}. Task 1, being a binary task, displays a notably higher inter-annotator agreement, indicating minimal ambiguity in the annotation process. In contrast, Task 2 exhibits a substantial decrease in agreement, primarily due to two significant factors.
\begin{equation} \label{eq:21}
    \kappa = \frac{P_o - P_e}{1 - P_e}
\end{equation}
First, Task 2 is dependent on Task 1, and second, the inherent complexity of Task 2 leads to various possible interpretations. Notably, Task 2 may involve cases where a single image corresponds to multiple labels, potentially causing discrepancies between annotators. While both tasks exhibit a moderate level of agreement, Task 2, which focuses on humanitarian images, shows significantly lower agreement compared to humanitarian text. The level of agreement and the reliable data percentage after the annotation process are shown in the~\autoref{Table5}. 
\begin{table}[pos=ht]
    \centering
    \caption{Level of Agreement and Reliability Percentage for Calculating Kappa Scores}\label{Table5}
  \begin{tabular}{p{0.65in}lcc}
   \toprule
    \textbf{Kappa Value} & \textbf{Level of Agreement} & \textbf{Reliable Data(\%)} \\
        \midrule
        0--0.20 & None & 0--4\% \\
        0.21--0.39 & Minimal & 4--15\% \\
        0.40--0.59 & Weak & 15--35\% \\
        0.60--0.79 & Moderate & 35--63\% \\
        0.80--0.90 & Strong & 64--81\% \\
        Above 0.90 & Almost Perfect & 82--100\% \\
          \bottomrule
    \end{tabular}
\end{table}
The inter-annotator agreement is quantified using Cohen's Kappa coefficient, a statistical measure that accounts for chance agreement, and it is computed as follows.
As shown in~\cref{eq:21}, $P_o$ is the actual agreement of the annotators. It is calculated by comparing the number of times annotators had the same annotations for given data.\\
$P_e$ is the probability of the two annotators agreeing on annotation calculated by multiplying the probability of each annotator predicting a class adds for all the classes. This is deducted from the total calculation to ensure that the agreement is not by chance. 
\subsection{Creation of Crisis Lexicons}
In the domain of crisis management, a well-constructed crisis lexicon list is essential, facilitating the clear and accurate communication of vital information. By integrating crisis-specific keywords in social media posts, stakeholders can efficiently sift through extensive online content, accessing pertinent information directly linked to the ongoing crisis. Olteanu \textit{et al.}~\citep{olteanu2014crisislex} formulated a lexicon list tailored to floods, encompassing 360 crisis-related terms. Additionally, we sourced 400 crisis lexicons from various online sources, such as Wikipedia\footnote{https://en.wikipedia.org/wiki/Earthquake}, and Disaster Words Vocabulary-List\footnote{https://www.vocabulary.com/lists/32961}. Our innovative approach involved leveraging a text encoder to generate embeddings for these meticulously curated crisis lexicons, along with word-level embeddings for tweets. By computing the cosine similarity between the crisis lexicon embeddings and tweet embeddings, we pinpointed words surpassing a threshold of 0.8, subsequently integrating them into the crisis lexicon list. These comprehensive crisis lexicon lists offer domain-specific insights, potentially enhancing the ability to discern pertinent content shared on social media platforms.

\subsection{Comparison Methods}
We explore a range of state-of-the-art unimodal and multimodal techniques. Among the unimodal methods, we delve into BERT and ResNet. For multimodal approaches, we examine state-of-the-art methods such as ViLBERT, VilT, and VisualBERT, along with SCBD, MAN, CAMM, RoBERTaMFT, FixMatchLS$_{\text{img+eda}}$, and DMCC.\\

\textbf{{BERT}:} BERT (Bidirectional Encoder Representations from Transformers) \citep{BERT} model is pre-trained on vast corpora, enabling it to capture contextual nuances and semantics, making it effective for vast NLP tasks, including question-answering, text classification, and language understanding.\\

\textbf{{ResNet}:} ResNet (Residual Network) \citep{RESnet} model addresses the issues of vanishing gradient in deep convolutional neural networks by introducing residual connections. ResNet has established itself as a cornerstone in the field of computer vision by achieving state-of-the-art performance in image recognition tasks.
\\

\textbf{{VisualBERT}:} VisualBERT \citep{visualBERT} is a groundbreaking model that combines vision and language understanding by integrating the Transformer architecture with both convolutional and recurrent neural networks. It leverages pre-trained BERT embeddings alongside visual features, enabling multimodal tasks such as visual question answering and image captioning.\\

\textbf{{VilBERT}:} Vision and Language BERT (VilBERT) \citep{vilbert} model is a pioneering architecture that integrates visual and textual information within a single BERT-based framework. Through pre-training on vast amounts of multimodal data, ViLBERT captures nuanced relationships between vision and language, enabling it to excel in tasks like visual question answering, image-text retrieval, and image captioning. \\

\textbf{{VilT}:} The VilT (Vision-and-Language Transformer) \citep{vilt} model harnesses the power of transformer architecture to seamlessly fuse textual and visual information. By jointly pre-trained on different vision and language data, VilT achieves remarkable capabilities in tasks like image captioning, visual question answering, and cross-modal retrieval, offering a versatile and powerful solution for bridging the gap between vision and language domains. \\

\textbf{{SCBD}:}  SCBD (SSE-Cross-BERT-DenseNet) \citep{CABD} model presents an approach to real-time event detection, particularly in emergency response scenarios. The model introduces a cross-attention module to filter out irrelevant components from text and image inputs, addressing the limitations of unimodal approaches. Additionally, SCBD employs a multimodal graph-based approach, allowing stochastic transitions between embeddings of different multimodal pairs during training to improve regularization and handle limited training data effectively. \\

\textbf{{MANN}:} Multimodal Adversarial Neural Network (MANN) \citep{MANN} addresses a significant challenge in disaster-related message detection on social media by introducing an approach to identifying posts related to emerging disaster events. It has the ability to extract comprehensive features from both text and images, enabling accurate detection of disaster-related content and adaptability to unforeseen events through adversarial training. \\

\textbf{{CAMM}:} The Cross-Attention Multi-Modal (CAMM) \citep{CAMM} deep neural network is proposed for classifying multimodal disaster data on social media. Integrating text and image information from Twitter posts using attention mechanisms, CAMM employs the attention mask from the textual modality to highlight relevant features in the visual modality. This cross-attention-based method emphasizes its efficacy in capturing pertinent cross-domain features in both text and image data.\\

\textbf{FixMatchLS$_{\text{img+eda}}$:} FixMatchLS{\text{\_img+eda}}~\citep{FixMatchLS} leverages the advantages of text and image modalities by integrating back translation and easy data augmentation techniques into the FixMatchLS model, an algorithm recognized for its semi-supervised learning capabilities. FixMatchLS$_{\text{img+eda}}$ builds upon the foundation of the MMBT (Multimodal Bitransformers) model. To fine-tune the FixMatchLS$_{\text{img+eda}}$ model, a dynamic weighting mechanism is introduced for the labeled and unlabeled loss components. \\

\textbf{RoBERTaMFT:} RoBERTa-based multimodal fusion transformer \citep{RoBERTaMFT}, addresses the challenges of multimodal learning. Its co-learning methodology effectively addresses issues related to recognition performance limitations and data imbalances among modalities. RoBERTaMFT is poised to offer significant advancements in multimodal understanding and classification by promoting balanced and effective learning across different data modalities.\\

\textbf{DMCC:} Deep Multimodal Crisis Categorization Framework (DMCC) \citep{DMCC} employs a two-level fusion strategy to seamlessly integrate textual and visual data. At the feature level, the MCA block ensures an effective fusion of these modalities. This holds the potential to enhance real-time crisis assessment and resource allocation, offering valuable support to emergency responders in dynamic and evolving situations.

\subsection{Evaluation Metrics} \label{EM}
 In our study, we assess a model's effectiveness using well-defined metrics such as accuracy, precision, recall, and F1-score. Accuracy offers a general measure of correctness. Precision evaluates the accuracy of positive predictions, while recall gauges the model's ability to correctly identify positive instances. The F1-Score, a harmonic mean of precision and recall, provides a balanced evaluation by considering both false positives and false negatives.

\subsection{ Experimental Results}
We first present the experimental setup, then present our experimental results and perform comparative assessments across various datasets. Initially, we compare the results of our proposed method with unimodal methods such as BERT \citep{BERT}, and ResNet \citep{RESnet}. Subsequently, we compare our proposed method with multimodal state-of-the-art methods, such as  VisualBERT \citep{visualBERT}, VilT \citep{vilt}, ViLBERT \citep{vilbert}. SCBD\citep{CABD}, MANN\citep{MANN}, CAMM\citep{CAMM}, FixMatchLS$_{\text{img+eda}}$ \citep{FixMatchLS}, RoBERTaMFT \citep{RoBERTaMFT}, and DMCC \citep{DMCC}. Furthermore, we conduct a comprehensive evaluation to determine the performance improvements associated with each modality and constituent component integrated into our proposed model.

\subsubsection{Experimental Setup}
In this study, all neural models are implemented using the PyTorch and TensorFlow frameworks, executed on a Linux server equipped with an Intel(R) Xeon(R) Silver 4215R CPU @3.20 GHz, 256GB of RAM, and a 16GB NVIDIA Tesla T4 GPU. The pre-trained BERT, CLIP, and ResNet models are accessible through the Hugging Face repository. We adopt the AdamW optimizer and set the learning rate hyper-parameter for the model to $5e-5$.
For the Feature Extraction and Interaction module (FEIM), we configure the dimensionality of hidden representations to $d = 768$ for the textual modality and $1024$ for the visual modality. Similarly, for the Multimodal Graph Learning module (MGLM), the dimensionality of hidden representations for both textual and visual modalities is set to $d = 512$. To mitigate overfitting, a dropout rate of $0.2$ is applied. The training process is conducted over 100 epochs.\\\\
We systematically explore hyperparameters for our proposed model, such as learning rate, weight decay, epochs, dropout, batch size, and optimizer, with the goal of identifying optimal configurations for improved training and overall performance. We conduct comprehensive experiments to determine the most effective hyperparameter values. For the learning rate and weight decay, we test a range from  $5e-3$ to  $5e-7$, with  $5e - 5$ identified as optimal. The number of epochs vary from $50$ to $120$, with $100$ epochs selected for optimal model convergence and performance enhancement. Batch sizes of $16, 32, 64$, and $128$ are evaluated, with a batch size of $32$ yielding the best results, balancing computational resources and memory constraints effectively. Exploration of dropout rates from $0.1$ to $0.5$ reveals $0.2$ as optimal, providing effective regularization without compromising learning capacity. Finally, after evaluating various optimizer options such as Adam, AdamW, and RMSProp, AdamW was chosen for its significant performance in optimizing the learning process. After rigorous experimentation, we determine the optimal hyperparameter configuration for our model, comprising a learning rate of $5e-5$, batch size of $32$, $100$ epochs, a dropout rate of $0.2$, and utilization of the AdamW optimizer. These selections aim to achieve the best balance in model performance.\\ 
The CrisisSpot network comprises several key components, each with its unique set of network parameters. Firstly, the Feature Extraction and Interaction Module (FEIM) employs one linear layer for projecting the textual features in the shared embedding space for the effective alignment of the dimensions as that of visual modality. We then employ 2 dense layers with Rectified Linear Unit (ReLU) activation functions for nonlinear transformations with output dimensions of 1024 each. Each layer is followed by a dropout layer of 0.2. The final concatenated vector, Fused Attended Vector (FAV), with dimensions of 2048, is passed to the multimodal adaptive network where it initially undergoes a hierarchical reduction in dimensionality. It passes through successive fully connected layers, achieving dimensions of 1024, 512, and 256, respectively. Each layer incorporates a dropout rate of 0.2. The refined output is then directed to the Joint Fusion Learning Network (JFLN) for further processing. The FEIM component plays a crucial role in extracting and interacting with features pertinent to crisis analysis.\\
Next, the Multimodal Graph Learning Module (MGLM) consists of two graph neural network layers, one for each modality. The output dimension of the graph neural network layers is set to 512. These layers are followed by a dropout rate of 0.2 after each layer. Following the concatenation process, the unified vector achieves a dimensionality of 1024. Subsequently, we utilize two consecutive fully connected layers: the first layer reduces the dimensionality to 512 with a dropout rate of 0.2, followed by a second layer, further reducing it to 256 with the same dropout rate. This processed vector is subsequently directed to the JFLN, enabling the seamless integration of information from diverse sources.\\
The Social Context Feature Module (SCFM) utilizes two dense layers with a ReLU activation function to capture social context features. The output dimension of the first layer is set to 21, the second layer reduces the dimensions to 16. These features incorporate relevant social context information into the predictions. \\
Lastly, the Multimodal Fusion Network Module (MFNM) integrates features from all the modules. We employ four dense layers with output dimensions of 528, 256, 128, and 64, respectively maintaining the dropout rate of 0.2 after each layer. The final layer employs a softmax activation function, with its output dimensions dependent on the number of classes being predicted. These selected network parameters ensure that the CrisisSpot network can effectively extract, process, and fuse information from multiple sources, enabling accurate crisis analysis and prediction.

\subsubsection{Performance Comparison on CrisisMMD} \label{PCC}
To affirm the effectiveness of CrisisSpot in addressing the Informative and Humanitarian Tasks on the CrisisMMD dataset \citep{CRISISMMD}, we carry out a thorough comparative analysis against unimodal and multimodal state-of-the-art approaches. We measure the performance of these methods using well-established metrics as discussed in \autoref{PCC}. We concentrate our comparison on methods adhering to the original data split provided by the CrisisNLP repository.
The results present in \autoref{Table6} demonstrate that CrisisSpot significantly outperforms both the unimodal and multimodal state-of-the-art methods. Our proposed approach yields impressive results, with an accuracy of 97.58\% and an F1-score of 98.23\% for the Informative Task. 
\begin{table}[pos=ht]
  \caption{Performance Comparison on CrisisMMD for Informative and Humanitarian Tasks}\label{Table6}
  \centering
  \begin{tabular}{lcccccccc}
    \toprule
    \multirow{2}{*}{\textbf{Method}} & \multicolumn{4}{c}{\textbf{Informative Task}} & \multicolumn{4}{c}{\textbf{Humanitarian Task}} \\
    \cmidrule(lr){2-5} \cmidrule(lr){6-9}
    & \textbf{Acc} & \textbf{Prec} & \textbf{Rec} & \textbf{F1} & \textbf{Acc} & \textbf{Prec} & \textbf{Rec} & \textbf{F1} \\
    \midrule
    BERT \citep{BERT} & 86.22 & 86.01 & 86.22 & 86.11 & 81.25 & 82.13 & 82.72 & 82.42 \\
    ResNet \citep{RESnet} & 84.88 & 84.76 & 84.89 & 84.82 & 81.33 & 81.29 & 81.98 & 81.63 \\
    Visual BERT \citep{visualBERT} & 88.10 & 86.29 & 87.23 & 86.76 & 84.05 & 82.22 & 83.56 & 82.88 \\
    VilT \citep{vilt} & 87.60 & 88.29 & 89.18 & 88.73 & 83.30 & 83.82 & 84.92 & 84.37 \\
    ViLBERT \citep{vilbert} & 88.25 & 88.92 & 90.14 & 89.53 & 84.33 & 84.95 & 86.52 & 85.73 \\
    SCBD \citep{CABD} & 88.52 & 91.13 & 91.84 & 91.48 & 80.62 & 80.77 & 80.52 & 80.64 \\
    MANN \citep{MANN} & 84.09 & 86.38 & 90.58 & 88.43 & 78.42 & 78.59 & 78.42 & 78.50 \\
    CAMM \citep{CAMM} & 87.22 &89.56  & 91.65  & 90.59 & 82.19 & 82.31  & 81.88 & 82.09  \\
    FixMatchLS$_{\text{img+eda}}$\citep{FixMatchLS} & - & 91.00 & 90.80 & 90.50 & - & 88.00 & 87.90 & 87.80 \\
    RoBERTaMFT \citep{RoBERTaMFT} & 88.90 & - & - & 87.40 & 84.90 & - & - & 80.40 \\
    DMCC \citep{DMCC} & 92.24 & 92.24 & 92.23 & 92.23 & 88.00 & 87.95 & 87.80 & 87.72 \\
    \midrule
    \textbf{CrisisSpot} & \textbf{97.58}& \textbf{96.70} & \textbf{99.80} & \textbf{98.23} & \textbf{90.01} & \textbf{90.87} & \textbf{90.04} & \textbf{90.13} \\
    \bottomrule
  \end{tabular}
\end{table}
As depicted in \autoref{Table6}, the performance of pre-trained unimodal methods, such as, BERT and ResNet is subpar as they are designed to process either text or image data, thus lacking the capability to harness multimodal information. VisualBERT \citep{visualBERT} underperforms due to its reliance on a single self-attention mechanism for handling both text and image data, potentially hindering its ability to capture intricate relationships between modalities. VilT \citep{vilt} faces challenges due to the absence of convolutional layers and reliance solely on transformers, impacting its ability to capture local features essential in computer vision tasks. ViLBERT inherits positional encoding from BERT, potentially limiting its effectiveness in capturing spatial relationships within images compared to convolutional layers. Additionally, ViLBERT's limited attention for cross-modality interactions may hinder its performance. SCBD \citep{CABD} utilizes SSE-Graph, which relies on the knowledge graph for swapping embeddings. However, it may not explicitly capture the complex structural relationships within the graph. Additionally, depending on the size of the knowledge graph and the frequency of swapping, SSE-Graph may face scalability challenges. Handling large-scale graphs with frequent embedding swaps might lead to increased computational complexity. CAMM \citep{CAMM} uses Bi-LSTMs to process text and VGG-16 for image feature extraction. It requires access to the entire input sequence before making predictions, which can be impractical for real-time tasks requiring low latency. MANN \citep{MANN} hinges on the quality of adversarial examples generated. If these examples aren't well-crafted representations of real-world challenges, the model might not learn robust features. DMCC \citep{DMCC} framework relies on the text modality's attention mask to highlight relevant areas in the image. This may lead to errors if the text doesn't perfectly correspond to the image content.\\
We attempt to incorporate various techniques to address the limitations of state-of-the-art methods. We attribute CrisisSpot's performance enhancement to the incorporation of graph-based learning, Social Context Features, and Inverted Dual Embedded Attention which enriches the feature representation of modalities. Our proposed attention mechanism, IDEA, enables CrisisSpot to capture both similar and dissimilar pairs more effectively, facilitating its ability to attend to essential patterns. Our proposed method captures both modality-specific features and cross-modal relationships, contributing to the improved performance of our model. Furthermore, the integration of Social Contextual Features enhance the model's context understanding by augmenting it with text-centric and user-centric scores. The various features such as User Informative Score (UIS), Crisis Informative Score (CIS), emotional context, and user engagement history enhance the classification outcomes. UIS gauges the proportion of informative content posted by a user compared to their total volume of posted content. A UIS score of 1 denotes a user has the polarity towards posting highly informative content, while 0 indicates the user's tendency to post non-informative content. This score is integrated into the classification model as a feature, adjusting the weight of each user's contributions during classification. Also, the Crisis Informative Score (CIS) aids in classification by quantifying the semantic attributes of the text, thereby facilitating the identification of crisis-related vocabulary within the text. Additionally, we leverage data on user engagement history to assess their influence. This includes factors such as friend count, reply count, number of likes, and other user features to further understand individual engagement patterns.\\
Also, it has been observed that emotional context significantly shapes classification outcomes. Through emotion and sentiment analysis techniques, we discern the emotional tone of content by analyzing sentiment polarity and identifying specific emotions expressed, including joy, sadness, anger, or fear. By incorporating sentiment and emotional cues, our classification outcomes are refined, enabling a more nuanced interpretation of user-generated content. The integration of these features forms the foundation of a robust classification framework.

\subsubsection{Performance Comparison of TSEqD}
To evaluate the effectiveness of CrisisSpot, we conduct a comparative analysis against existing unimodal and multimodal state-of-the-art benchmarks using our TSEqD dataset. CrisisSpot's performance is assessed against a variety of methods, both unimodal and multimodal, to comprehensively gauge its efficacy in disaster content classification. Despite their widespread use, unimodal methods, such as, BERT and ResNet suffer from inherent limitations. BERT, originally designed for textual data, lacks the ability to process visual information, while ResNet, a convolutional neural network (CNN), operates solely on images, overlooking textual cues crucial for disaster content understanding. Multimodal methods such as ViLBERT, VisualBERT, and VILT introduce their own set of challenges. ViLBERT's attention mechanism may not adequately handle cross-modal interactions, impacting performance. VisualBERT, though promising in its integration of both textual and visual information, faces limitations due to its reliance on a single self-attention mechanism for processing both modalities, potentially leading to suboptimal feature extraction. VILT, exclusively reliant on transformer architectures, lacks convolutional layers essential for capturing local image features, potentially limiting its ability to discern fine-grained visual details.
\begin{table}[pos=ht]
  \caption{Performance Comparison on TSEqD for Informative and Humanitarian Tasks}\label{Table7}
  \centering
  \begin{tabular}{lcccccccc}
    \toprule
    \textbf{Method} & \multicolumn{4}{c}{\textbf{Informative Task}} & \multicolumn{4}{c}{\textbf{Humanitarian Task}} \\
    \cmidrule(lr){2-5} \cmidrule(lr){6-9}
    & \textbf{Acc} & \textbf{Prec} & \textbf{Rec} & \textbf{F1} & \textbf{Acc} & \textbf{Prec} & \textbf{Rec} & \textbf{F1} \\
    \midrule
    BERT \citep{BERT} & 76.42 & 78.29 & 79.35 & 78.81 & 72.22 & 74.44 & 75.84 & 75.13 \\
    ResNet \citep{RESnet} & 77.85 & 79.54 & 80.23 & 79.89 & 72.85 & 75.54 & 76.23 & 75.88 \\
    Visual BERT \citep{visualBERT} & 78.56 & 80.21 & 81.28 & 80.74 & 74.56 & 76.21 & 78.28 & 77.23 \\
    VilT \citep{vilt} & 78.92 & 79.65 & 81.28 & 80.45 & 74.02 & 75.23 & 78.25 & 76.71 \\
    ViLBERT \citep{vilbert} & 79.99 & 82.24 & 86.86 & 84.48 & 76.00 & 78.27 & 80.19 & 79.22 \\
    SCBD \citep{CABD} & 80.79 & 83.16 & 89.58 & 86.25 & 77.50 & 78.03 & 78.58 & 78.30 \\
    MANN \citep{MANN} & 80.14 & 82.76 & 88.55 & 85.55 & 76.80 & 77.25 & 78.95 & 78.09 \\
    CAMM \citep{CAMM} & 81.08 & 83.89 & 90.34 & 86.99 & 78.00 & 79.43 & 80.34 & 79.88 \\
    FixMatchLS$_{\text{img+eda}}$ \citep{FixMatchLS} & 80.23 & 82.95 & 89.96 & 86.31 & 75.23 & 74.95 & 78.96 & 77.01 \\
    RoBERTaMFT \citep{RoBERTaMFT} & 78.23 & 82.55 & 88.25 & 85.30 & 74.23 & 74.55 & 78.25 & 76.36 \\
    DMCC \citep{DMCC} & 81.10 & 84.15 & 90.13 & 87.03 & 78.10 & 77.15 & 78.80 & 77.96 \\
    \midrule
    \textbf{CrisisSpot} & \textbf{84.42} & \textbf{84.57} & \textbf{96.20} & \textbf{90.01} & \textbf{80.55} & \textbf{81.87} & \textbf{82.20} & \textbf{82.03} \\
    \bottomrule
  \end{tabular}
\end{table}
The results, as presented in \autoref{Table7}, showcase CrisisSpot's impressive performance metrics. For the Informative Task, CrisisSpot achieves an accuracy of 84.42\%, a precision of 84.57\%, a recall of 96.20\%, and an F1-score of 90.01\%. These metrics demonstrate significant improvements over both unimodal and multimodal state-of-the-art benchmarks, highlighting CrisisSpot's superiority in disaster content classification. In comparison to RoBERTaMFT, CrisisSpot outperforms it by 8.68\% in accuracy and 10.83\% in the F1-score. Similarly, when benchmarked against DMCC, CrisisSpot exhibits performance gains of 3.32\% in accuracy, 0.42\% in precision, 6.07\% in recall, and 2.98\% in F1-score. These results emphasize CrisisSpot's consistent superiority over other methods across various evaluation metrics. Furthermore, CrisisSpot is compared against SCBD \citep{CABD}, MANN \citep{MANN}, and CAMM \citep{CAMM}, all of which represent multimodal state-of-the-art methods. Notably, CrisisSpot outperforms these methods by substantial margins across all evaluation metrics. This comprehensive comparison underscores the robustness and effectiveness of CrisisSpot in the context of disaster content classification. In conclusion, the superior performance exhibited by CrisisSpot reaffirms its position as an advanced solution for disaster content classification. By surpassing existing benchmarks, CrisisSpot demonstrates its ability to leverage both unimodal and multimodal information effectively, thereby enhancing decision-making processes in crisis scenarios.

\subsubsection{Performance Gain Analysis}
We assess the performance improvement achieved by CrisisSpot through a systematic analysis on CrisisMMD for Informative tasks. Initially, we evaluate our model's performance with different combinations of modalities and subsequently explore the impact of various attention mechanisms. We evaluate the effect of incorporating different combinations of model components. Finally, we analyze various language, vision, and graph methods. 
\begin{table}[pos=ht]
\caption{Performance Gain Analysis of Different Model Variants}
\label{Table9}
\centering
\begin{tabular}{{lcccccc}}
\toprule
\textbf{Caption} & \textbf{Model} & \textbf{Acc} & \textbf{Pre} & \textbf{Rec} & \textbf{F1} \\
\midrule
\multirow{4}{*}{Modality-based Comparison}
& CrisisSpot (w/o Image) & 92.56 & 93.84 & 95.35 & 94.14 \\
& CrisisSpot (w/o Text) & 91.72 & 92.20 & 94.66 & 93.41 \\ 
& CrisisSpot (w/o SCF) & 94.63 & 95.86 & 96.42 & 96.14 \\
& CrisisSpot (T+I+SCF) & 97.58 & 96.70 & 99.80 & 98.23 \\
\midrule
\multirow{4}{*}{Attention-based Comparison} & No Attention & 95.86 & 94.32 & 98.72 & 96.47 \\
 & Self Attention & 95.95 & 94.48 & 98.85 & 96.58 \\
 & Cross Attention & 96.12 & 95.26 & 98.97 & 97.21 \\
 & IDEA & 97.58 & 96.70 & 99.80 & 98.23 \\
\midrule
\multirow{4}{*}{Social Context Feature-based Comparison} & CrisisSpot(w/o UIS) & 96.08 & 95.23 & 98.31 & 96.74 \\
 & CrisisSpot(w/o CIS) & 97.03 & 96.18 & 99.21 & 97.67 \\
 & CrisisSpot(w/o OF) & 96.63 & 95.75 & 98.85 & 97.28 \\
 & CrisisSpot(U+C+OF) & 97.58 & 96.70 & 99.80 & 98.23 \\
\midrule
\multirow{4}{*}{Model Component Comparison} & CrisisSpot(w/o SCFM) & 94.63 & 95.86 & 96.42 & 96.14 \\
 & CrisisSpot(w/o MGLM) & 93.98 & 94.20 & 95.98 & 94.08 \\
 & CrisisSpot(w/o FEIM) & 95.32 & 95.44 & 97.26 & 96.34 \\
 & CrisisSpot(SF+ML+FI) & 97.58 & 96.70 & 99.80 & 98.23 \\
\bottomrule
\end{tabular}
\end{table}
\paragraph{Analyzing Combinations of Modalities}
\mbox{}\\
We systematically evaluate the performance of CrisisSpot across a range of modalities combinations. Given the diverse characteristics of each modality, it is crucial to examine the impact of different modality combinations employed in our approach. Our method integrates inputs from three distinct modalities: textual (T), visual (I), and Social Context Features (SCF). The performance assessments of these various modality combinations are elaborated in~\autoref{Table9}.\\
To evaluate the importance of image input, we exclude it while preserving text and Social Context Features, resulting in CrisisSpot (w/o Images). However, this exclusion may impose limitations on the model. Firstly, the absence of images results in a loss of crucial visual context, impairing the model's ability to comprehensively understand the data. Additionally, without integrating images, the model misses out on the synergistic benefits derived from combining text and visual data. This leads to a restricted understanding of the data, hindering its performance. Furthermore, the absence of images prevents the utilization of attention mechanisms, limiting the model's ability to focus on relevant content. The lack of cross-modal feature interaction further restricts the model's effectiveness. Consequently, CrisisSpot (w/o Images) may face challenges in achieving optimal performance compared to the proposed model that incorporates all modalities.\\
The CrisisSpot (w/o Text) approach, designed to assess the significance of textual data, excludes text while retaining images and Social Context Features as inputs. By excluding text, we aim to analyze the impact of textual information on the model's performance. This enables us to discern the specific role of text in enhancing the understanding and classification of crisis-related content. CrisisSpot (w/o Text) presents several technical implications. Firstly, without textual information, the model may struggle to grasp the semantic context embedded within crisis-related content, potentially leading to misinterpretations or incomplete understanding. Additionally, text often provides valuable contextual cues crucial for accurately categorizing crisis-related data. The absence of text input deprives the method of this contextual information, reducing its ability to make informed decisions or classifications.  Without text modality, the model may struggle to capture cross-modal relationships effectively, limiting its ability to leverage complementary information from both modalities. Consequently, the exclusion of text may result in decreased classification accuracy and a loss of discriminative power, as textual data often contributes discriminative features essential for accurate classification. In summary, CrisisSpot (w/o Text) may experience reduced semantic understanding, limited contextual information, inability to capture text-image relationships, and decreased classification performance due to the absence of textual data. These technical implications underscore the critical importance of incorporating textual information into multimodal methods for effective crisis content classification.\\
CrisisSpot (w/o SCF) focuses on evaluating the significance of Social Context Features (SCF) by excluding them while preserving textual and visual inputs. This configuration enables an assessment of the individual contributions of SCF to CrisisSpot's overall performance. SCF comprises five factors: User Information Source (UIS), Crisis Information Source (CIS), user engagement history, emotional analysis, and sentiment analysis. CrisisSpot (w/o SCF) would lack crucial information regarding the social context surrounding crisis-related content. Specifically, without incorporating factors such as User Information Source (UIS), Crisis Information Source (CIS), user engagement history, emotional analysis, and sentiment analysis, the method would miss out on valuable insights into the credibility, relevance, and sentiment of the content. This omission could result in several consequences. Firstly, the CrisisSpot would have a limited understanding of the social context surrounding crisis events, potentially leading to misinterpretations or incomplete assessments of the severity and urgency of the situation. Additionally, without considering factors such as, user engagement history and sentiment analysis, the method may struggle to accurately gauge the relevance and impact of crisis-related content, affecting its ability to prioritize and respond effectively. Moreover, the absence of social context scores may hinder the model's ability to make informed decisions regarding the credibility and trustworthiness of information. Furthermore, without user-centric scores and other social context features, CrisisSpot (w/o SCF) may overlook valuable opportunities for engaging with affected communities or leveraging user-generated content to enhance situational awareness and response efforts. In summary, omitting social context scores from CrisisSpot (w/o SCF) would result in a diminished ability to accurately assess the social context surrounding crisis events, potentially compromising the effectiveness of the model in disaster management tasks.\\
CrisisSpot (T+I+SCF), integrating textual modality, visual modality, and Social Context Features (SCF), demonstrates superior performance compared to all other variants across various evaluation metrics as can seen in \autoref{Table9}. This enhancement in CrisisSpot (T+I+SCF) compared to its variants stems from the effective utilization of multimodal information, cross-modal relationships, attention mechanisms, and Social Context Features inherent in the posts. These results underscore the efficacy of our proposed multimodal approach and highlight the importance of leveraging each modality integrated into the model for achieving superior performance in crisis content classification tasks.

\paragraph{Analysis of Attention Mechanisms}
\mbox{}\\
We conduct a comprehensive analysis of different attention mechanisms in CrisisSpot, ranging from methods without attention to those employing classical attention mechanisms such as additive attention, self-attention, and cross-attention.
All methods are trained and evaluated on the CrisisMMD dataset using four key metrics: accuracy, precision, recall, and F1-score. We gain crucial insights into the performance of these attention mechanisms. Our findings reveal that compared to the proposed method without attention, all classical attention mechanisms exhibit notable enhancement. This highlights the significance of directing the model's focus toward relevant information, thereby enhancing CrisisSpot's ability to identify crisis effectively.
However, the most remarkable performance is demonstrated by our proposed IDEA (Inverted Dual Embedding Attention) model. CrisisSpot with IDEA outperforms all compared methods across all metrics, showcasing a remarkable recall of 99.80\%. This suggests that IDEA excels in capturing a higher proportion of actual crisis content, due to its utilization of inverted attention to filter out irrelevant information.
Moreover, IDEA maintains a high precision of 96.70\%, indicating its proficiency in minimizing false positives. This robust performance underscores the effectiveness of IDEA in accurately identifying crisis data.
In summary, our ablation study confirms the vital role of attention mechanisms in CrisisSpot for disaster content classification. It underscores the substantial contribution of the proposed IDEA mechanism, which leverages inverted attention and dual embedding spaces to achieve superior performance compared to classical attention mechanisms. These findings provide compelling evidence for the value of IDEA in enhancing CrisisSpot's performance and its potential to achieve a more accurate and comprehensive understanding of crisis content.

\paragraph{Analysis of Social Context Features}
\mbox{}\\
In this section, we delve into the intricate role of Social Context Features (SCF) within CrisisSpot, focusing on User Informative Score (UIS), Crisis Informative Score (CIS), and other related features as depicted from \autoref{Table9}. These features play a pivotal role in understanding the context and informativeness of social media content during the crisis. CrisisSpot (w/o UIS) may struggle to discern the credibility and significance of user-contributed content during a crisis. UIS serves as a crucial indicator of content informativeness, allowing the model to prioritize credible and informative content over noise and misinformation. Without UIS, CrisisSpot's ability to distinguish between reliable and less informative content may be compromised, potentially leading to an increase in false positives and reduced overall accuracy in crisis detection.
Additionally, the absence of a Crisis Informative Score poses a significant challenge in identifying crisis-related vocabulary within textual data. CIS serves as a key indicator of the presence of crisis-relevant terms, enabling CrisisSpot to discern the informativeness of crisis-related content. CrisisSpot (w/o CIS) may struggle to capture essential crisis signals embedded within textual data, leading to potential misclassifications or oversights in crisis classification. As a result, the model's overall precision, recall, and F1-score may suffer, impacting its effectiveness in crisis content classification.
In \autoref{Table9}, for simplicity, we denote CrisisSpot (w/o other features) as CrisisSpot (w/o OF). Neglecting additional features such as Sentiment Quotient, Emotion Quotient, and User Engagement Metrics deprives CrisisSpot of valuable contextual information necessary for understanding the emotional context and relevance of crisis-related content. Emotion and sentiment analysis are crucial for interpreting the underlying sentiments and nuances expressed in crisis-related posts, while user engagement metrics provide insights into the relevance and impact of content within the crisis context. Without considering these features, CrisisSpot may overlook important interactions, emotional cues, and contextual nuances present in crisis-related content, potentially 
\begin{figure*}[pos=ht]
    \centering
    \includegraphics[width=\linewidth]{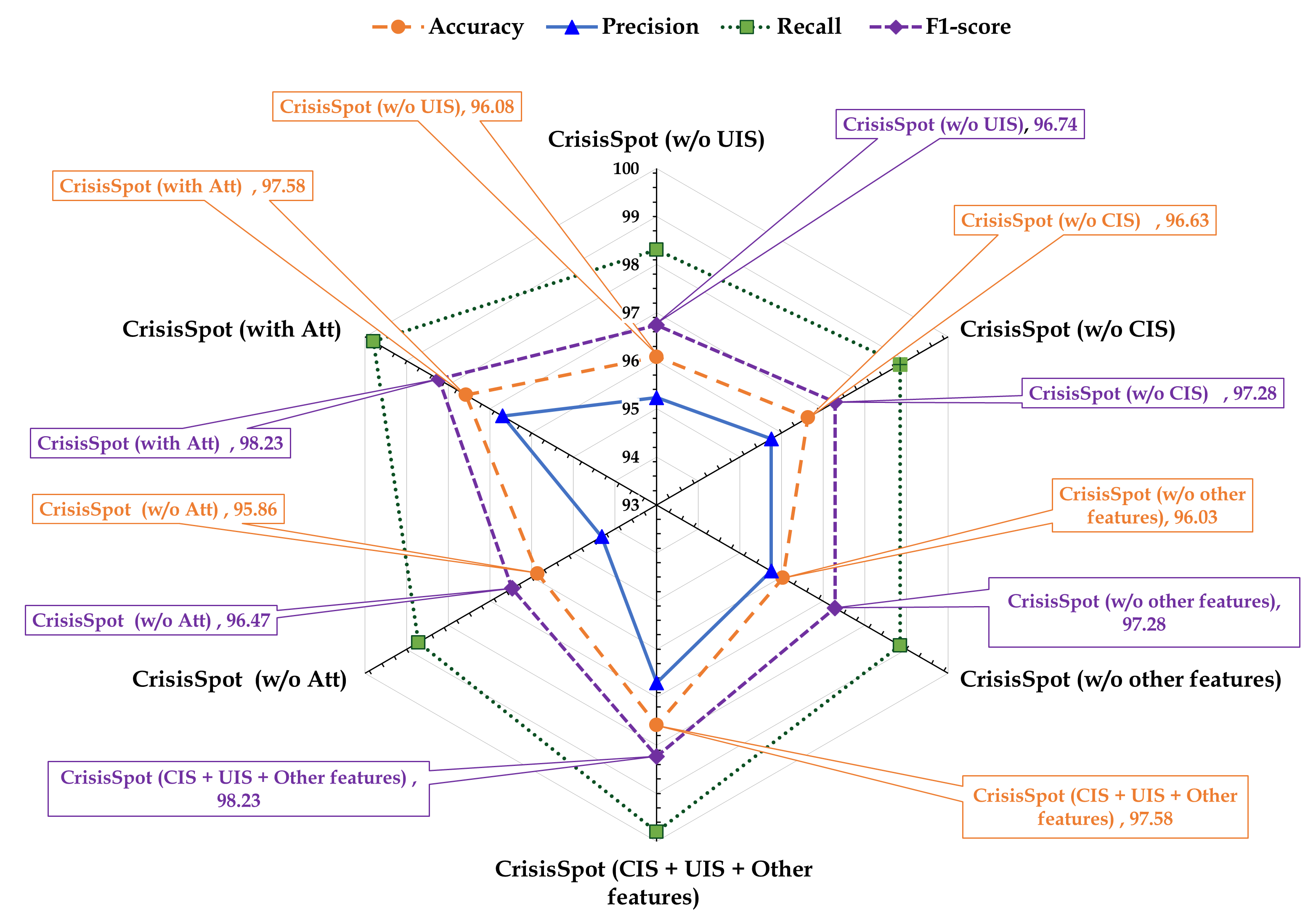}
    \caption{Performance Gain Analysis on Social Context Features and Attention Mechanism}
    \label{Fig.4}
\end{figure*}
leading to suboptimal performance in crisis classification. Our proposed, CrisisSpot (CIS + UIS + Other Features), integrates all SCF components, yielding superior results across all evaluation metrics. By incorporating UIS, CIS, and other related features, CrisisSpot achieves a comprehensive understanding of crisis-related content, enabling more accurate and reliable crisis classification. The inclusion of these features enhances CrisisSpot's ability to discern credible, informative, and contextually relevant content amidst the noise of social media during crises, underscoring the significance of SCF integration in crisis management and response.
\paragraph{Model Component Analysis}
\mbox{}\\
In this section, we delve deeply into the intricate components of our CrisisSpot model, dissecting their individual contributions and collective impact on performance, as illustrated in \autoref{Table9}. Through a series of ablation experiments, we systematically explore the role of each module, shedding light on their significance in crisis content classification. The various variants are as:\\
Our first variant, CrisisSpot (w/o SCFM), excludes the SCFM, focusing solely on the FEIM and MGLM modules. Despite the absence of SCFM, which typically provides invaluable contextual insights, this configuration remarkably maintains competitive performance metrics. The effective interaction between textual and visual features facilitated by FEIM, coupled with knowledge propagation through MGLM, underscores the model's ability to capture nuanced patterns within crisis-related content. However, without SCFM, the model may lack deep contextual understanding derived from social context features, potentially leading to suboptimal performance in scenarios where such information is critical.\\
Next, we evaluate the impact of the MGLM, CrisisSpot (w/o MGLM), by excluding it from the model architecture, while retaining SCFM and FEIM. Despite the omission of MGLM, known for enriching feature representation through graph knowledge propagation, our model exhibits significant performance. The contextual understanding provided by SCFM, coupled with the effective interaction between textual and visual modalities facilitated by FEIM, underscores the model's resilience in crisis content classification. However, the absence of MGLM may limit the model's ability to exploit complex relationships between different modalities, potentially resulting in a less nuanced understanding of crisis-related content.\\
The next variant, CrisisSpot (w/o FEIM), removes the FEIM module from the model, depriving it of attention mechanisms and modality-specific features. Despite this limitation, the model demonstrates commendable performance, highlighting the efficacy of multimodal integration and contextual understanding provided by SCFM and MGLM. Leveraging features extracted from the multimodal encoder, along with text-centric and user-centric scores, the model compensates for the absence of FEIM, showcasing competitive accuracy, precision, recall, and F1-score. However, without FEIM, the model may struggle to capture fine-grained patterns and relationships within and between modalities, potentially leading to reduced performance in scenarios requiring a nuanced understanding of crisis-related content.\\
The proposed, CrisisSpot (FEIM + MGLM + SCFM), method integrates all components, synergistically leveraging attention mechanisms, modality-specific features, and SCFs to achieve superior performance. By combining FEIM, MGLM, and SCFM, CrisisSpot surpasses other variants, demonstrating the effectiveness of our holistic approach. The seamless integration of these modules enables the model to capture nuanced patterns within crisis-related content, resulting in significant accuracy, precision, recall, and F1-score. Here, the absence of any individual component may significantly diminish the model's overall performance, underscoring the importance of comprehensive multimodal integration and SCFs in crisis content classification. It is worth noting that in \cref{Fig.4}, CrisisSpot (w/o Att) refers to No Attention and CrisisSpot (with Attn) refers to the proposed method with IDEA attention mechanism.
In summary, our model component analysis provides a nuanced understanding of CrisisSpot's architecture, highlighting the critical role of each module in enhancing performance. Through systematic experimentation and insightful observations, we underscore the efficacy of our proposed approach in crisis content classification.

\begin{figure}
        \centering
        \includegraphics[width=1\linewidth]{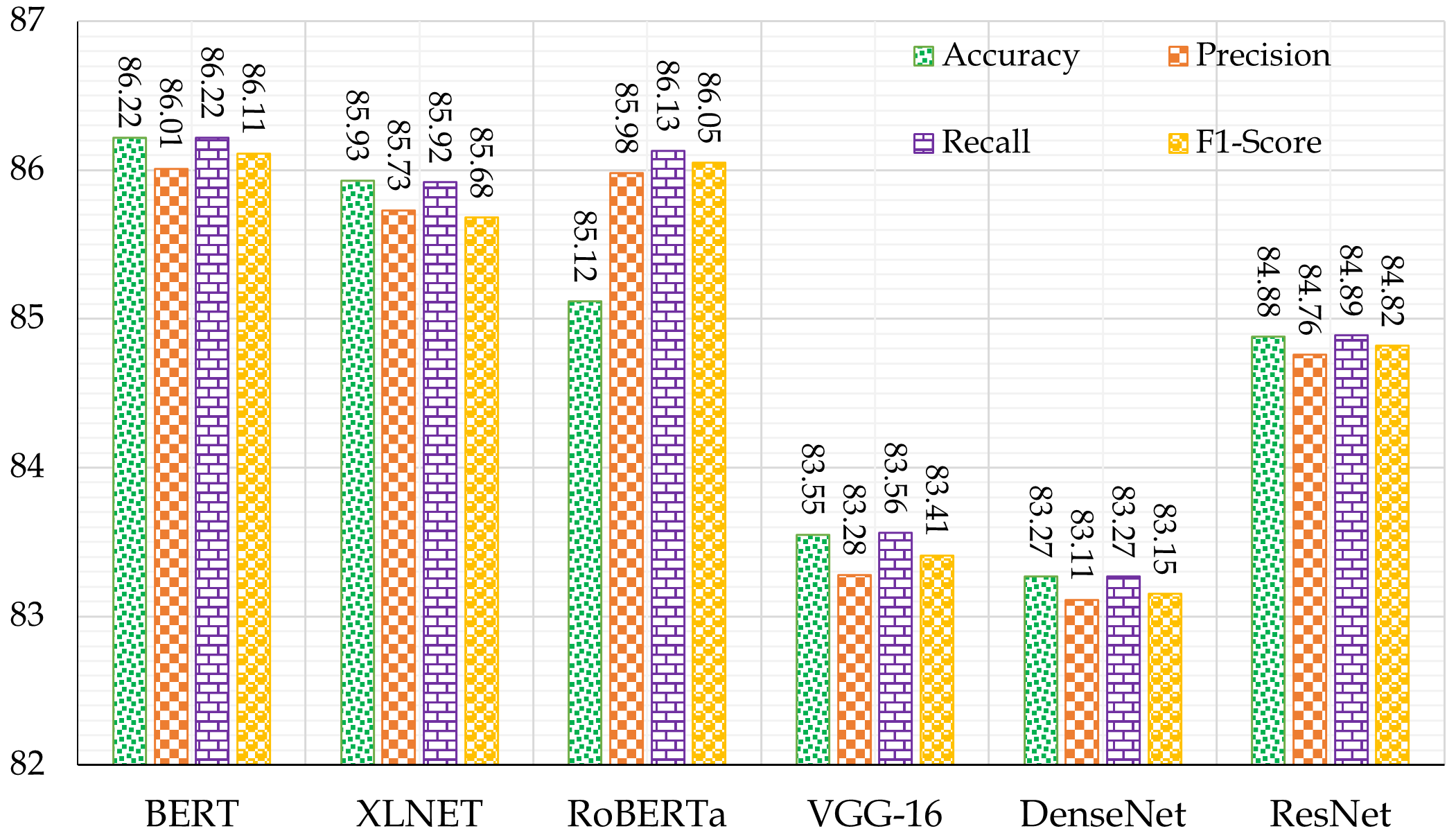}
        \caption{Analysis of Language and Vision Methods}
        \label{LVG}
    \end{figure}

\paragraph{Analysis of Language, Vision and Graph Methods}
\mbox{}\\
In this section, we conduct a comprehensive analysis of language, vision, and graph models, focusing on their effectiveness in addressing crisis content classification tasks. Our selection of methods includes prominent representatives from each domain, namely BERT, ResNet, and GraphSAGE, each chosen for their distinct strengths and widespread adoption in the respective fields of natural language processing, computer vision, and graph representation learning, respectively.\\
BERT has demonstrated remarkable capability in capturing semantic nuances and contextual dependencies within textual data. Its superior performance across various natural language processing tasks, such as text classification, stems from its ability to leverage bidirectional context information effectively. \autoref{LVG} illustrates the accuracy, precision, recall, and F1-score achieved by various language and vision methods. Our analysis reveals that BERT outperforms both XLNet and RoBERTa across all the evaluation parameters. 
Similarly, ResNet is preferred for its depth and effectiveness in image classification, tackling vanishing gradient issues with skip connections and residual blocks. Its hierarchical feature capture across scales makes it a top choice for visual representation learning. ResNet when benchmarked against DenseNet and  VGG-16 outperforms both the models with significant margins as clearly visible from \autoref{LVG}. 
Furthermore, GraphSAGE was compared with GCN, in which GraphSAGE outperformed GCN across all evaluation metrics as depicted in \autoref{Tables11}. GraphSAGE is a pioneering method in graph representation learning that addresses the challenge of learning node embeddings in heterogeneous graph structures. By aggregating information from local neighborhoods through graph convolutions, 
GraphSAGE generates informative embeddings that encapsulate the structural and semantic properties of nodes in the graph. Its scalability, adaptability, and robustness make GraphSAGE a compelling choice for classification tasks. Our selection of BERT, ResNet, and GraphSAGE is driven by their proven effectiveness and widespread adoption in their respective domains. 
\begin{table}[pos=ht]
  \caption{Analysis of Graph Methods}
  \label{Tables11}
  \centering
  \begin{tabular}{lcccc}
    \toprule
    \textbf{Method} & \textbf{Accuracy} & \textbf{Precision} & \textbf{Recall} & \textbf{F1}  \\
    \midrule
    GCN & 87.05 & 87.60 & 87.66 & 87.63 \\
    GraphSAGE & 88.12 & 88.77 & 88.80 & 88.78 \\
    \bottomrule
  \end{tabular}
\end{table}
Numerical evidence, as depicted in the \autoref{LVG} and \autoref{Tables11}, further corroborates the superior performance of these baseline models compared to alternative methods across various classification benchmarks. CrisisSpot outperforms all of these methods across all evaluation metrics.

\section{Discussion} \label{Section.5}
This article presents CrisisSpot, a novel framework for classifying disaster-related content. CrisisSpot incorporates Social Context Features (SCFs) encompassing factors such as user post-informativeness, crisis post-informativeness, emotional context, and user engagement history, providing a multifaceted view of disaster-related data. Our proposed method introduces the Inverted Dual Embedded Attention (IDEA) mechanism, which adeptly directs attention to both aligned and misaligned information patterns across diverse modalities. Additionally, CrisisSpot employs a graph-based approach designed for the reconstruction of features which helps in enhancing the classification outcomes. Our proposed method leverages both cross-modal relationships and discriminative features to get a richer representation of the data. By leveraging these approaches, our work not only fills the gap in utilizing SCF but also enhances the model’s capability to effectively handle disaster-related information, thereby contributing significantly to the field of disaster content classification.
\subsection{Advantages over Existing Works}\label{sec:advantages}
Our proposed multimodal attentive framework offers several key advantages that elevate state-of-the-art methods for enhancing disaster content classification:
\begin{itemize}
    \item {Integration of Social Context Features (SCF):} A crucial aspect often neglected in prior works is the integration of SCF with the textual content. Our framework addresses this gap by introducing Social Context Features (SCF). SCF incorporates user-centric and text-centric information through metrics such as User Informative Score (UIS), Crisis Informative Score (CIS), and User Engagement metrics. UIS assesses the content informativeness based on their past posting history, while CIS quantifies the semantic relevance of the text content to the specific disaster event. These features empower the model to prioritize informative content posted by the users and content that is highly relevant to the crisis at hand.
    \item {Attention Mechanism}: The proposed attention mechanism, Inverted Dual Embedded Attention (IDEA), focuses on both aligned and misaligned information patterns, potentially improving the model's ability to handle complex multimodal data compared to conventional attention mechanisms limited to aligned information.
   \item Graph-based Representation Learning:  Many existing methods struggle to capture the intricate relationships between disaster-related content. Our framework addresses this limitation by integrating a graph-based approach. CrisisSpot utilizes cosine similarity and updates nodes using information from the entire dataset, leading to potentially more accurate feature refinement compared to existing methods limited by intra-cluster information or suboptimal distance metrics. CrisisSpot excels at learning local graph topology, crucial for capturing complex relationships, unlike knowledge graph-based methods that may face scalability challenges with large datasets.
    \item Improved Classification Performance through Multimodal Learning: Most of the existing disaster content classification methods often rely solely on textual data. Our framework leverages a multimodal attentive approach, incorporating information from various modalities such as text and images. This fusion of modalities leads to a richer representation of disaster content, ultimately resulting in superior classification performance.
\end{itemize}
\subsection{Limitations and Future Work}
CrisisSpot demonstrates strong performance in disaster-related content classification. However, ongoing development can enhance its capabilities and broaden its applicability. The limitations of CrisisSpot are as follows:
\begin{itemize}
    \item Monolingual Focus: CrisisSpot currently focuses on monolingual data, achieving high accuracy within specific languages. This limits its effectiveness in multilingual disaster scenarios. Future work should explore techniques for incorporating multi-lingual capabilities. This could involve leveraging pre-trained multilingual transformer models or employing language identification modules to handle code-switching and mixed-language content.
    \item Social Context Feature (SCF) Dependence: The inclusion of SCFs significantly enhances CrisisSpot's classification performance. However, situations where SCF data might be limited necessitate exploring alternative feature extraction techniques. 
    \item Data Size and Augmentation: The size and quality of training data significantly impact performance. Exploring data augmentation techniques, such as back-translation, synonym substitution, or paraphrasing, could be a valuable avenue for further research. Data augmentation can potentially improve the model's robustness by exposing it to a wider range of variations in disaster-related content.
\end{itemize}
\subsection{Computational Efficiency and Scalability}
CrisisSpot achieves significant advancements in disaster-related content classification while maintaining efficiency and scalability. This section details these aspects of our proposed approach.\\\\
\textbf{Model Complexity and Efficiency:} One key strategy to ensure computational efficiency is CrisisSpot's utilization of three non-trainable encoders for feature extraction. These pre-trained encoders eliminate the need to train a significantly larger model from scratch, resulting in a total of 10.275 million trainable parameters. This parameter count falls within the range of typical deep learning architectures, mitigating potential computational burdens during training. CrisisSpot incorporates Social Context Features (SCF) that require minimal configuration, with only 21 neurons needed for SCF training. The proposed method demonstrates a real-time inference speed of 8.46 microseconds per sample.\\\\
\textbf{Modular Architecture and Time Complexity:} CrisisSpot leverages a modular architecture for efficient computation. The time complexity of each module is given as follows: Feature Extraction and Interaction: This module utilizes some operations such as scaled dot-product attention and a few matrix multiplications, contributing an overall time complexity of \( O(mn^2) \), where \( m \) represents the maximum sequence length, and \( n \) signifies the feature dimensionality of both textual and visual modalities. Multimodal Graph Learning Module: In this module, the complexity hinges on neighbor sampling and layer stacking, resulting in time complexity of \( O(r^kn'd^2) \), where \( r \) represents the fixed number of neighbors sampled per node, \( k \) signifies the number of layers, \( n' \) represents the number of nodes, and \( d \) represents the feature dimensions of both textual and visual modalities, which are fixed. Social Context Feature Module: This module operates in \( O(i*j) \), where \( i \) represents the input dimension and \( j \) signifies the output dimension. However, since both \( i \) and \( j \) are small, the impact of this time complexity on the overall computation is minimal. The  Multimodal Fusion Network comprises three networks with a combined eleven dense layers having a time complexity of $O(i'*j')$ for each sample where $i'$ and $j'$ denote the fixed input and output dimensions, respectively. This time complexity is minimal compared to other modules. The overall time complexity can be expressed as the sum of individual time complexities \( O(mn^2 + r^kn'd^2) \).\\\\
\textbf{Scalability:}
It is also crucial to acknowledge that CrisisSpot, similar to other deep learning methods, employs a mini-batch training technique. This technique allows the model to process data in smaller batches during training. This approach offers two key advantages for scalability: \\
\begin{itemize}
    \item {Efficient Handling of Varying Data Sizes:} Mini-batch training enables CrisisSpot to efficiently process data across different scales, from smaller datasets to large-scale. Regardless of the volume of data, the model can be trained effectively while maintaining computational efficiency.
    \item { Flexibility for Distributed Training:} Mini-batch training readily adapts to distributed computing environments. By distributing the training load across multiple machines or GPUs, CrisisSpot can be scaled to handle even larger datasets efficiently. CrisisSpot's design is inherently scalable. The underlying transformer architecture can be readily parallelized on GPUs or TPUs for handling large datasets.
\end{itemize}
CrisisSpot prioritizes computational efficiency and scalability through its use of non-trainable encoders, real-time inference speed, and the adoption of mini-batch training. These design choices position CrisisSpot as a solution for disaster response, where accurate information is critical.
\section{Conclusion} \label{Section.6}
Our research explores the multifaceted role of social media in crisis situations, acknowledging both its potential for real-time information dissemination and the associated challenges regarding information veracity. To address these complexities, we propose CrisisSpot, a novel multimodal attentive learning framework designed to classify disaster-related content on social media platforms. CrisisSpot harnesses the robust capabilities of transformer-based methods, integrating Social Context Features such as User Informative Score (UIS) and crisis Informative Score (CIS), alongside emotional context, and user engagement metrics. UIS evaluates the proportion of informative content posted by a user, while CIS quantifies crisis-related vocabulary within the content. Additionally, leveraging user engagement histories such as friend count, reply count, number of likes, and other user features helps the model to understand individual engagement patterns, thereby refining classification outcomes. Through sentiment and emotion analysis, CrisisSpot discerns emotional tones and specific emotions expressed in content, facilitating a nuanced interpretation of user-generated posts. These integrations empower CrisisSpot to assess content informativeness, emotional cues, and user posting history, thereby improving classification performance. Furthermore, CrisisSpot's effectiveness hinges on these innovations: the Inverted Dual Embedded Attention (IDEA) mechanism and a graph-based approach for feature reconstruction. IDEA excels at identifying both aligned and misaligned information patterns within multimodal content, leading to more accurate disaster classification compared to methods that solely rely on supportive information. Additionally, the graph-based approach iteratively aggregates information from neighboring nodes, enhancing the feature representation. Through rigorous evaluation of both CrisisMMD and TSEqD datasets, CrisisSpot demonstrates superior performance compared to existing techniques across various key performance indicators (KPIs) such as accuracy, precision, recall, and F1-score for both Informative and Humanitarian tasks. Notably, for the Informative task on CrisisMMD, CrisisSpot surpasses DMCC by a substantial margin, achieving a 5.34\% improvement in accuracy and a 5.99\% improvement in the F1-score. CrisisSpot demonstrates clear advantages over augmentation-based techniques such as FixMatchLS$_{\text{img+eda}}$, achieving a 5.70\% improvement in accuracy and a 7.73\% improvement in F1-score. Most significantly, CrisisSpot outperforms RoBERTaMFT, a robust transformer-based method, by 8.68\% in accuracy and a staggering 10.83\% in F1-score. These results highlight the effectiveness of CrisisSpot's design choices, particularly the incorporation of Social Context Features, IDEA mechanism, and graph-based feature reconstruction. This paves the way for significant advancements in disaster response, ultimately saving lives during emergencies by enabling more efficient and accurate classification of critical information on social media platforms. Furthermore, the development of the TSEqD dataset offers a valuable resource for the research community, fostering further exploration in disaster-related content classification. CrisisSpot serves as a powerful method, leveraging the power of social media data to improve disaster response and public safety efforts during emergencies.
\bibliographystyle{elsarticle-harv}
\bibliography{export1}

\begin{thebibliography}{63}
\expandafter\ifx\csname natexlab\endcsname\relax\def\natexlab#1{#1}\fi
\providecommand{\url}[1]{\texttt{#1}}
\providecommand{\href}[2]{#2}
\providecommand{\path}[1]{#1}
\providecommand{\DOIprefix}{doi:}
\providecommand{\ArXivprefix}{arXiv:}
\providecommand{\URLprefix}{URL: }
\providecommand{\Pubmedprefix}{pmid:}
\providecommand{\doi}[1]{\href{http://dx.doi.org/#1}{\path{#1}}}
\providecommand{\Pubmed}[1]{\href{pmid:#1}{\path{#1}}}
\providecommand{\bibinfo}[2]{#2}
\ifx\xfnm\relax \def\xfnm[#1]{\unskip,\space#1}\fi
\bibitem[{Abavisani et~al.(2020)Abavisani, Wu, Hu, Tetreault and Jaimes}]{CABD}
\bibinfo{author}{Abavisani, M.}, \bibinfo{author}{Wu, L.}, \bibinfo{author}{Hu, S.}, \bibinfo{author}{Tetreault, J.}, \bibinfo{author}{Jaimes, A.}, \bibinfo{year}{2020}.
\newblock \bibinfo{title}{Multimodal categorization of crisis events in social media}, in: \bibinfo{booktitle}{Proceedings of the IEEE/CVF Conference on Computer Vision and Pattern Recognition}, pp. \bibinfo{pages}{14679--14689}.
\newblock \DOIprefix\doi{10.1109/CVPR42600.2020.01469}.
\bibitem[{Agarwal et~al.(2020)Agarwal, Leekha, Sawhney and Shah}]{CrisisDIAS}
\bibinfo{author}{Agarwal, M.}, \bibinfo{author}{Leekha, M.}, \bibinfo{author}{Sawhney, R.}, \bibinfo{author}{Shah, R.R.}, \bibinfo{year}{2020}.
\newblock \bibinfo{title}{Crisis-dias: Towards multimodal damage analysis - deployment, challenges and assessment}.
\newblock \bibinfo{journal}{AAAI 2020 - 34th AAAI Conference on Artificial Intelligence} \DOIprefix\doi{10.1609/aaai.v34i01.5369}.
\bibitem[{Ahmad et~al.(2022)Ahmad, Jindal, Mukuntha, Ekbal and Bhattachharyya}]{zishan}
\bibinfo{author}{Ahmad, Z.}, \bibinfo{author}{Jindal, R.}, \bibinfo{author}{Mukuntha, N.S.}, \bibinfo{author}{Ekbal, A.}, \bibinfo{author}{Bhattachharyya, P.}, \bibinfo{year}{2022}.
\newblock \bibinfo{title}{Multi-modality helps in crisis management: An attention-based deep learning approach of leveraging text for image classification}.
\newblock \bibinfo{journal}{Expert Systems with Applications} \bibinfo{volume}{195}.
\newblock \DOIprefix\doi{10.1016/j.eswa.2022.116626}.
\bibitem[{Alam et~al.(2023)Alam, Alam, Hasan, Hasnat, Imran and Ofli}]{alam2023medic}
\bibinfo{author}{Alam, F.}, \bibinfo{author}{Alam, T.}, \bibinfo{author}{Hasan, M.A.}, \bibinfo{author}{Hasnat, A.}, \bibinfo{author}{Imran, M.}, \bibinfo{author}{Ofli, F.}, \bibinfo{year}{2023}.
\newblock \bibinfo{title}{Medic: a multi-task learning dataset for disaster image classification}.
\newblock \bibinfo{journal}{Neural Computing and Applications} \bibinfo{volume}{35}.
\newblock \DOIprefix\doi{10.1007/s00521-022-07717-0}.
\bibitem[{Alam et~al.(2018a)Alam, Joty and Imran}]{Alam2018}
\bibinfo{author}{Alam, F.}, \bibinfo{author}{Joty, S.}, \bibinfo{author}{Imran, M.}, \bibinfo{year}{2018}a.
\newblock \bibinfo{title}{Graph based semi-supervised learning with convolution neural networks to classify crisis related tweets}.
\newblock \bibinfo{journal}{12th International AAAI Conference on Web and Social Media, ICWSM 2018} \DOIprefix\doi{10.1609/icwsm.v12i1.15047}.
\bibitem[{Alam et~al.(2018b)Alam, Ofli and Imran}]{CRISISMMD}
\bibinfo{author}{Alam, F.}, \bibinfo{author}{Ofli, F.}, \bibinfo{author}{Imran, M.}, \bibinfo{year}{2018}b.
\newblock \bibinfo{title}{Crisismmd: Multimodal twitter datasets from natural disasters}.
\newblock \bibinfo{journal}{12th International AAAI Conference on Web and Social Media, ICWSM 2018} \DOIprefix\doi{10.1609/icwsm.v12i1.14983}.
\bibitem[{Alam et~al.(2020)Alam, Ofli, Imran, Alam and Qazi}]{alam}
\bibinfo{author}{Alam, F.}, \bibinfo{author}{Ofli, F.}, \bibinfo{author}{Imran, M.}, \bibinfo{author}{Alam, T.}, \bibinfo{author}{Qazi, U.}, \bibinfo{year}{2020}.
\newblock \bibinfo{title}{Deep learning benchmarks and datasets for social media image classification for disaster response}.
\newblock \bibinfo{journal}{Proceedings of the 2020 IEEE/ACM International Conference on Advances in Social Networks Analysis and Mining, ASONAM 2020} \DOIprefix\doi{10.1109/ASONAM49781.2020.9381294}.
\bibitem[{Anshul et~al.(2023)Anshul, Pranav, Rehman and Kumar}]{Anshul}
\bibinfo{author}{Anshul, A.}, \bibinfo{author}{Pranav, G.S.}, \bibinfo{author}{Rehman, M.Z.U.}, \bibinfo{author}{Kumar, N.}, \bibinfo{year}{2023}.
\newblock \bibinfo{title}{A multimodal framework for depression detection during covid-19 via harvesting social media}.
\newblock \bibinfo{journal}{IEEE Transactions on Computational Social Systems} \DOIprefix\doi{10.1109/TCSS.2023.3309229}.
\bibitem[{Bahdanau et~al.(2014)Bahdanau, Cho and Bengio}]{additive_attention}
\bibinfo{author}{Bahdanau, D.}, \bibinfo{author}{Cho, K.}, \bibinfo{author}{Bengio, Y.}, \bibinfo{year}{2014}.
\newblock \bibinfo{title}{Neural machine translation by jointly learning to align and translate}.
\newblock \bibinfo{journal}{arXiv preprint arXiv:1409.0473} .
\bibitem[{Baltrusaitis et~al.(2019)Baltrusaitis, Ahuja and Morency}]{MMLSURVEY}
\bibinfo{author}{Baltrusaitis, T.}, \bibinfo{author}{Ahuja, C.}, \bibinfo{author}{Morency, L.P.}, \bibinfo{year}{2019}.
\newblock \bibinfo{title}{Multimodal machine learning: A survey and taxonomy}.
\newblock \DOIprefix\doi{10.1109/TPAMI.2018.2798607}.
\bibitem[{Basu et~al.(2019)Basu, Shandilya, Khosla, Ghosh and Ghosh}]{8715653}
\bibinfo{author}{Basu, M.}, \bibinfo{author}{Shandilya, A.}, \bibinfo{author}{Khosla, P.}, \bibinfo{author}{Ghosh, K.}, \bibinfo{author}{Ghosh, S.}, \bibinfo{year}{2019}.
\newblock \bibinfo{title}{Extracting resource needs and availabilities from microblogs for aiding post-disaster relief operations}.
\newblock \bibinfo{journal}{IEEE Transactions on Computational Social Systems} \bibinfo{volume}{6}.
\newblock \DOIprefix\doi{10.1109/TCSS.2019.2914179}.
\bibitem[{Biggers et~al.(2023)Biggers, Mohanty and Manda}]{SMD}
\bibinfo{author}{Biggers, F.B.}, \bibinfo{author}{Mohanty, S.D.}, \bibinfo{author}{Manda, P.}, \bibinfo{year}{2023}.
\newblock \bibinfo{title}{A deep semantic matching approach for identifying relevant messages for social media analysis}.
\newblock \bibinfo{journal}{Scientific Reports} \bibinfo{volume}{13}.
\newblock \DOIprefix\doi{10.1038/s41598-023-38761-y}.
\bibitem[{Chaudhuri and Bose(2020)}]{chaudhuri2020}
\bibinfo{author}{Chaudhuri, N.}, \bibinfo{author}{Bose, I.}, \bibinfo{year}{2020}.
\newblock \bibinfo{title}{Exploring the role of deep neural networks for post-disaster decision support}.
\newblock \bibinfo{journal}{Decision Support Systems} \bibinfo{volume}{130}.
\newblock \DOIprefix\doi{10.1016/j.dss.2019.113234}.
\bibitem[{hin Cheung and man Lam(2022)}]{negative}
\bibinfo{author}{hin Cheung, T.}, \bibinfo{author}{man Lam, K.}, \bibinfo{year}{2022}.
\newblock \bibinfo{title}{Crossmodal bipolar attention for multimodal classification on social media}.
\newblock \bibinfo{journal}{Neurocomputing} \bibinfo{volume}{514}.
\newblock \DOIprefix\doi{10.1016/j.neucom.2022.09.140}.
\bibitem[{Devlin et~al.(2019)Devlin, Chang, Lee and Toutanova}]{BERT}
\bibinfo{author}{Devlin, J.}, \bibinfo{author}{Chang, M.W.}, \bibinfo{author}{Lee, K.}, \bibinfo{author}{Toutanova, K.}, \bibinfo{year}{2019}.
\newblock \bibinfo{title}{Bert: Pre-training of deep bidirectional transformers for language understanding}.
\newblock \bibinfo{journal}{NAACL HLT 2019 - 2019 Conference of the North American Chapter of the Association for Computational Linguistics: Human Language Technologies - Proceedings of the Conference} \bibinfo{volume}{1}.
\bibitem[{Esposito et~al.(2022)Esposito, Galli, Moscato and Sperlí}]{credibility}
\bibinfo{author}{Esposito, C.}, \bibinfo{author}{Galli, A.}, \bibinfo{author}{Moscato, V.}, \bibinfo{author}{Sperlí, G.}, \bibinfo{year}{2022}.
\newblock \bibinfo{title}{Multi-criteria assessment of user trust in social reviewing systems with subjective logic fusion}.
\newblock \bibinfo{journal}{Information Fusion} \bibinfo{volume}{77}.
\newblock \DOIprefix\doi{10.1016/j.inffus.2021.07.012}.
\bibitem[{Gao et~al.(2020)Gao, Li, Zhu and Wang}]{MANN}
\bibinfo{author}{Gao, W.}, \bibinfo{author}{Li, L.}, \bibinfo{author}{Zhu, X.}, \bibinfo{author}{Wang, Y.}, \bibinfo{year}{2020}.
\newblock \bibinfo{title}{Detecting disaster-related tweets via multimodal adversarial neural network}.
\newblock \bibinfo{journal}{IEEE Multimedia} \bibinfo{volume}{27}.
\newblock \DOIprefix\doi{10.1109/MMUL.2020.3012675}.
\bibitem[{Ghafarian and Yazdi(2020)}]{ghafarian2020}
\bibinfo{author}{Ghafarian, S.H.}, \bibinfo{author}{Yazdi, H.S.}, \bibinfo{year}{2020}.
\newblock \bibinfo{title}{Identifying crisis-related informative tweets using learning on distributions}.
\newblock \bibinfo{journal}{Information Processing and Management} \bibinfo{volume}{57}.
\newblock \DOIprefix\doi{10.1016/j.ipm.2019.102145}.
\bibitem[{Hamilton et~al.(2017)Hamilton, Ying and Leskovec}]{GraphSage}
\bibinfo{author}{Hamilton, W.L.}, \bibinfo{author}{Ying, R.}, \bibinfo{author}{Leskovec, J.}, \bibinfo{year}{2017}.
\newblock \bibinfo{title}{Inductive representation learning on large graphs}.
\newblock \bibinfo{journal}{Advances in Neural Information Processing Systems} \bibinfo{volume}{2017-December}.
\bibitem[{Hao and Wang(2020)}]{multimodal1}
\bibinfo{author}{Hao, H.}, \bibinfo{author}{Wang, Y.}, \bibinfo{year}{2020}.
\newblock \bibinfo{title}{Leveraging multimodal social media data for rapid disaster damage assessment}.
\newblock \bibinfo{journal}{International Journal of Disaster Risk Reduction} \bibinfo{volume}{51}.
\newblock \DOIprefix\doi{10.1016/j.ijdrr.2020.101760}.
\bibitem[{Hao et~al.(2017)Hao, Zhang, Liu, He, Liu, Wu and Zhao}]{cross}
\bibinfo{author}{Hao, Y.}, \bibinfo{author}{Zhang, Y.}, \bibinfo{author}{Liu, K.}, \bibinfo{author}{He, S.}, \bibinfo{author}{Liu, Z.}, \bibinfo{author}{Wu, H.}, \bibinfo{author}{Zhao, J.}, \bibinfo{year}{2017}.
\newblock \bibinfo{title}{An end-to-end model for question answering over knowledge base with cross-attention combining global knowledge}, in: \bibinfo{booktitle}{Proceedings of the 55th Annual Meeting of the Association for Computational Linguistics (Volume 1: Long Papers)}, pp. \bibinfo{pages}{221--231}.
\newblock \DOIprefix\doi{10.18653/v1/P17-1021}.
\bibitem[{He et~al.(2016)He, Zhang, Ren and Sun}]{RESnet}
\bibinfo{author}{He, K.}, \bibinfo{author}{Zhang, X.}, \bibinfo{author}{Ren, S.}, \bibinfo{author}{Sun, J.}, \bibinfo{year}{2016}.
\newblock \bibinfo{title}{Deep residual learning for image recognition}.
\newblock \bibinfo{journal}{Proceedings of the IEEE Computer Society Conference on Computer Vision and Pattern Recognition} \bibinfo{volume}{2016-December}.
\newblock \DOIprefix\doi{10.1109/CVPR.2016.90}.
\bibitem[{Hinton et~al.(2015)Hinton, Vinyals and Dean}]{Temp2}
\bibinfo{author}{Hinton, G.}, \bibinfo{author}{Vinyals, O.}, \bibinfo{author}{Dean, J.}, \bibinfo{year}{2015}.
\newblock \bibinfo{title}{Distilling the knowledge in a neural network}.
\newblock \bibinfo{journal}{arXiv preprint arXiv:1503.02531} .
\bibitem[{Hutto and Gilbert(2014)}]{VADER}
\bibinfo{author}{Hutto, C.J.}, \bibinfo{author}{Gilbert, E.}, \bibinfo{year}{2014}.
\newblock \bibinfo{title}{Vader: A parsimonious rule-based model for sentiment analysis of social media text}.
\newblock \bibinfo{journal}{Proceedings of the 8th International Conference on Weblogs and Social Media, ICWSM 2014} \DOIprefix\doi{10.1609/icwsm.v8i1.14550}.
\bibitem[{Imran et~al.(2015)Imran, Castillo, Diaz and Vieweg}]{10.1145/2771588}
\bibinfo{author}{Imran, M.}, \bibinfo{author}{Castillo, C.}, \bibinfo{author}{Diaz, F.}, \bibinfo{author}{Vieweg, S.}, \bibinfo{year}{2015}.
\newblock \bibinfo{title}{Processing social media messages in mass emergency: A survey}.
\newblock \bibinfo{journal}{ACM Computing Surveys} \bibinfo{volume}{47}.
\newblock \DOIprefix\doi{10.1145/2771588}.
\bibitem[{Imran et~al.(2020)Imran, Ofli, Caragea and Torralba}]{multimodal}
\bibinfo{author}{Imran, M.}, \bibinfo{author}{Ofli, F.}, \bibinfo{author}{Caragea, D.}, \bibinfo{author}{Torralba, A.}, \bibinfo{year}{2020}.
\newblock \bibinfo{title}{Using ai and social media multimodal content for disaster response and management: Opportunities, challenges, and future directions}.
\newblock \DOIprefix\doi{10.1016/j.ipm.2020.102261}.
\bibitem[{Jiang et~al.(2020)Jiang, Wang, Liu and Ling}]{attention}
\bibinfo{author}{Jiang, T.}, \bibinfo{author}{Wang, J.}, \bibinfo{author}{Liu, Z.}, \bibinfo{author}{Ling, Y.}, \bibinfo{year}{2020}.
\newblock \bibinfo{title}{Fusion-extraction network for multimodal sentiment analysis}.
\newblock \bibinfo{journal}{Lecture Notes in Computer Science (including subseries Lecture Notes in Artificial Intelligence and Lecture Notes in Bioinformatics)} \bibinfo{volume}{12085 LNAI}.
\newblock \DOIprefix\doi{10.1007/978-3-030-47436-2_59}.
\bibitem[{Khattar and Quadri(2022)}]{CAMM}
\bibinfo{author}{Khattar, A.}, \bibinfo{author}{Quadri, S.M.}, \bibinfo{year}{2022}.
\newblock \bibinfo{title}{Camm: Cross-attention multimodal classification of disaster-related tweets}.
\newblock \bibinfo{journal}{IEEE Access} \bibinfo{volume}{10}.
\newblock \DOIprefix\doi{10.1109/ACCESS.2022.3202976}.
\bibitem[{Kim et~al.(2021)Kim, Son and Kim}]{vilt}
\bibinfo{author}{Kim, W.}, \bibinfo{author}{Son, B.}, \bibinfo{author}{Kim, I.}, \bibinfo{year}{2021}.
\newblock \bibinfo{title}{Vilt: Vision-and-language transformer without convolution or region supervision}, in: \bibinfo{booktitle}{International Conference on Machine Learning}, \bibinfo{organization}{PMLR}. pp. \bibinfo{pages}{5583--5594}.
\bibitem[{Kohler et~al.(2018)Kohler, Daneshmand, Lucchi, Hofmann, Zhou and Neymeyr}]{BN2}
\bibinfo{author}{Kohler, J.}, \bibinfo{author}{Daneshmand, H.}, \bibinfo{author}{Lucchi, A.}, \bibinfo{author}{Hofmann, T.}, \bibinfo{author}{Zhou, M.}, \bibinfo{author}{Neymeyr, K.}, \bibinfo{year}{2018}.
\newblock \bibinfo{title}{Towards a theoretical understanding of batch normalization}.
\newblock \bibinfo{journal}{arXiv} .
\bibitem[{Koshy and Elango(2023)}]{koshy2023multimodal}
\bibinfo{author}{Koshy, R.}, \bibinfo{author}{Elango, S.}, \bibinfo{year}{2023}.
\newblock \bibinfo{title}{Multimodal tweet classification in disaster response systems using transformer-based bidirectional attention model}.
\newblock \bibinfo{journal}{Neural Computing and Applications} \bibinfo{volume}{35}.
\newblock \DOIprefix\doi{10.1007/s00521-022-07790-5}.
\bibitem[{Kumar et~al.(2019)Kumar, Singh and Saumya}]{kumar}
\bibinfo{author}{Kumar, A.}, \bibinfo{author}{Singh, J.P.}, \bibinfo{author}{Saumya, S.}, \bibinfo{year}{2019}.
\newblock \bibinfo{title}{A comparative analysis of machine learning techniques for disaster-related tweet classification}.
\newblock \bibinfo{journal}{IEEE Region 10 Humanitarian Technology Conference, R10-HTC} \bibinfo{volume}{2019-November}.
\newblock \DOIprefix\doi{10.1109/R10-HTC47129.2019.9042443}.
\bibitem[{Li et~al.(2020)Li, Zhang, Wang, Zhang, Wang, Gao, Duan, Tsoi and Wang}]{SocialMediaBenefits}
\bibinfo{author}{Li, L.}, \bibinfo{author}{Zhang, Q.}, \bibinfo{author}{Wang, X.}, \bibinfo{author}{Zhang, J.}, \bibinfo{author}{Wang, T.}, \bibinfo{author}{Gao, T.L.}, \bibinfo{author}{Duan, W.}, \bibinfo{author}{Tsoi, K.K.F.}, \bibinfo{author}{Wang, F.Y.}, \bibinfo{year}{2020}.
\newblock \bibinfo{title}{Characterizing the propagation of situational information in social media during covid-19 epidemic: A case study on weibo}.
\newblock \bibinfo{journal}{IEEE Transactions on Computational Social Systems} \bibinfo{volume}{7}.
\newblock \DOIprefix\doi{10.1109/TCSS.2020.2980007}.
\bibitem[{Li et~al.(2019)Li, Yatskar, Yin, Hsieh and Chang}]{visualBERT}
\bibinfo{author}{Li, L.H.}, \bibinfo{author}{Yatskar, M.}, \bibinfo{author}{Yin, D.}, \bibinfo{author}{Hsieh, C.J.}, \bibinfo{author}{Chang, K.W.}, \bibinfo{year}{2019}.
\newblock \bibinfo{title}{Visualbert: A simple and performant baseline for vision and language}.
\newblock \bibinfo{journal}{arXiv e-prints} , \bibinfo{pages}{arXiv--1908}.
\bibitem[{Lu et~al.(2019)Lu, Batra, Parikh and Lee}]{vilbert}
\bibinfo{author}{Lu, J.}, \bibinfo{author}{Batra, D.}, \bibinfo{author}{Parikh, D.}, \bibinfo{author}{Lee, S.}, \bibinfo{year}{2019}.
\newblock \bibinfo{title}{Vilbert: Pretraining task-agnostic visiolinguistic representations for vision-and-language tasks}, in: \bibinfo{booktitle}{Proceedings of the 33rd International Conference on Neural Information Processing Systems}, pp. \bibinfo{pages}{13--23}.
\bibitem[{Lu et~al.(2016)Lu, Yang, Batra and Parikh}]{hierarchical}
\bibinfo{author}{Lu, J.}, \bibinfo{author}{Yang, J.}, \bibinfo{author}{Batra, D.}, \bibinfo{author}{Parikh, D.}, \bibinfo{year}{2016}.
\newblock \bibinfo{title}{Hierarchical question-image co-attention for visual question answering}.
\newblock \bibinfo{journal}{Advances in Neural Information Processing Systems} , \bibinfo{pages}{289--297}.
\bibitem[{Luo et~al.(2019)Luo, Wang, Shao and Peng}]{BN1}
\bibinfo{author}{Luo, P.}, \bibinfo{author}{Wang, X.}, \bibinfo{author}{Shao, W.}, \bibinfo{author}{Peng, Z.}, \bibinfo{year}{2019}.
\newblock \bibinfo{title}{Towards understanding regularization in batch normalization}.
\newblock \bibinfo{journal}{7th International Conference on Learning Representations, ICLR 2019} .
\bibitem[{Madichetty and M.(2021)}]{madichetty2021}
\bibinfo{author}{Madichetty, S.}, \bibinfo{author}{M., S.}, \bibinfo{year}{2021}.
\newblock \bibinfo{title}{A novel method for identifying the damage assessment tweets during disaster}.
\newblock \bibinfo{journal}{Future Generation Computer Systems} \bibinfo{volume}{116}.
\newblock \DOIprefix\doi{10.1016/j.future.2020.10.037}.
\bibitem[{Madichetty et~al.(2023)Madichetty, M and Madisetty}]{madichetty2023roberta}
\bibinfo{author}{Madichetty, S.}, \bibinfo{author}{M, S.}, \bibinfo{author}{Madisetty, S.}, \bibinfo{year}{2023}.
\newblock \bibinfo{title}{A roberta based model for identifying the multi-modal informative tweets during disaster}.
\newblock \bibinfo{journal}{Multimedia Tools and Applications} \DOIprefix\doi{10.1007/s11042-023-14780-9}.
\bibitem[{Madichetty and Sridevi(2021)}]{Previous_work1}
\bibinfo{author}{Madichetty, S.}, \bibinfo{author}{Sridevi, M.}, \bibinfo{year}{2021}.
\newblock \bibinfo{title}{A neural-based approach for detecting the situational information from twitter during disaster}.
\newblock \bibinfo{journal}{IEEE Transactions on Computational Social Systems} \bibinfo{volume}{8}.
\newblock \DOIprefix\doi{10.1109/TCSS.2021.3064299}.
\bibitem[{Mohammad and Turney(2013)}]{Emolex}
\bibinfo{author}{Mohammad, S.M.}, \bibinfo{author}{Turney, P.D.}, \bibinfo{year}{2013}.
\newblock \bibinfo{title}{Crowdsourcing a word-emotion association lexicon}.
\newblock \bibinfo{journal}{Computational Intelligence} \bibinfo{volume}{29}.
\newblock \DOIprefix\doi{10.1111/j.1467-8640.2012.00460.x}.
\bibitem[{Mozafari et~al.(2019)Mozafari, Gomes, Leão and Gagné}]{Temp}
\bibinfo{author}{Mozafari, A.S.}, \bibinfo{author}{Gomes, H.S.}, \bibinfo{author}{Leão, W.}, \bibinfo{author}{Gagné, C.}, \bibinfo{year}{2019}.
\newblock \bibinfo{title}{Unsupervised temperature scaling: Post-processing unsupervised calibration of deep models decisions}.
\newblock \bibinfo{journal}{arXiv} \bibinfo{volume}{1905.00174}.
\bibitem[{Nguyen et~al.(2017)Nguyen, Ofli, Imran and Mitra}]{nguyen2017}
\bibinfo{author}{Nguyen, D.T.}, \bibinfo{author}{Ofli, F.}, \bibinfo{author}{Imran, M.}, \bibinfo{author}{Mitra, P.}, \bibinfo{year}{2017}.
\newblock \bibinfo{title}{Damage assessment from social media imagery data during disasters}.
\newblock \bibinfo{journal}{Proceedings of the 2017 IEEE/ACM International Conference on Advances in Social Networks Analysis and Mining, ASONAM 2017} \DOIprefix\doi{10.1145/3110025.3110109}.
\bibitem[{Nguyen et~al.(2020)Nguyen, Pernkopf and Kosmider}]{Temp3}
\bibinfo{author}{Nguyen, T.}, \bibinfo{author}{Pernkopf, F.}, \bibinfo{author}{Kosmider, M.}, \bibinfo{year}{2020}.
\newblock \bibinfo{title}{Acoustic scene classification for mismatched recording devices using heated-up softmax and spectrum correction}, in: \bibinfo{booktitle}{ICASSP 2020-2020 IEEE International Conference on Acoustics, Speech and Signal Processing (ICASSP)}, \bibinfo{organization}{IEEE}. pp. \bibinfo{pages}{126--130}.
\bibitem[{Olteanu et~al.(2014)Olteanu, Castillo, Diaz and Vieweg}]{olteanu2014crisislex}
\bibinfo{author}{Olteanu, A.}, \bibinfo{author}{Castillo, C.}, \bibinfo{author}{Diaz, F.}, \bibinfo{author}{Vieweg, S.}, \bibinfo{year}{2014}.
\newblock \bibinfo{title}{Crisislex: A lexicon for collecting and filtering microblogged communications in crises}.
\newblock \bibinfo{journal}{Proceedings of the 8th International Conference on Weblogs and Social Media, ICWSM 2014} \DOIprefix\doi{10.1609/icwsm.v8i1.14538}.
\bibitem[{Radford et~al.(2021)Radford, Kim, Hallacy, Ramesh, Goh, Agarwal, Sastry, Askell, Mishkin, Clark, Krueger and Sutskever}]{CLIP}
\bibinfo{author}{Radford, A.}, \bibinfo{author}{Kim, J.W.}, \bibinfo{author}{Hallacy, C.}, \bibinfo{author}{Ramesh, A.}, \bibinfo{author}{Goh, G.}, \bibinfo{author}{Agarwal, S.}, \bibinfo{author}{Sastry, G.}, \bibinfo{author}{Askell, A.}, \bibinfo{author}{Mishkin, P.}, \bibinfo{author}{Clark, J.}, \bibinfo{author}{Krueger, G.}, \bibinfo{author}{Sutskever, I.}, \bibinfo{year}{2021}.
\newblock \bibinfo{title}{Learning transferable visual models from natural language supervision}.
\newblock \bibinfo{journal}{Proceedings of Machine Learning Research} \bibinfo{volume}{139}.
\bibitem[{Rehman et~al.(2023)Rehman, Mehta, Singh, Kaushik and Kumar}]{SCF}
\bibinfo{author}{Rehman, M.Z.U.}, \bibinfo{author}{Mehta, S.}, \bibinfo{author}{Singh, K.}, \bibinfo{author}{Kaushik, K.}, \bibinfo{author}{Kumar, N.}, \bibinfo{year}{2023}.
\newblock \bibinfo{title}{User-aware multilingual abusive content detection in social media}.
\newblock \bibinfo{journal}{Information Processing and Management} \bibinfo{volume}{60}.
\newblock \DOIprefix\doi{10.1016/j.ipm.2023.103450}.
\bibitem[{Rezk et~al.(2023)Rezk, Elmadany, Hamad and Badran}]{DMCC}
\bibinfo{author}{Rezk, M.}, \bibinfo{author}{Elmadany, N.}, \bibinfo{author}{Hamad, R.K.}, \bibinfo{author}{Badran, E.F.}, \bibinfo{year}{2023}.
\newblock \bibinfo{title}{Categorizing crises from social media feeds via multimodal channel attention}.
\newblock \bibinfo{journal}{IEEE Access} \bibinfo{volume}{11}.
\newblock \DOIprefix\doi{10.1109/ACCESS.2023.3294474}.
\bibitem[{Rudra et~al.(2018)Rudra, Ganguly, Goyal and Ghosh}]{rudra2018}
\bibinfo{author}{Rudra, K.}, \bibinfo{author}{Ganguly, N.}, \bibinfo{author}{Goyal, P.}, \bibinfo{author}{Ghosh, S.}, \bibinfo{year}{2018}.
\newblock \bibinfo{title}{Extracting and summarizing situational information from the twitter social media during disasters}.
\newblock \bibinfo{journal}{ACM Transactions on the Web} \bibinfo{volume}{12}.
\newblock \DOIprefix\doi{10.1145/3178541}.
\bibitem[{Shaw et~al.(2018)Shaw, Uszkoreit and Vaswani}]{additive}
\bibinfo{author}{Shaw, P.}, \bibinfo{author}{Uszkoreit, J.}, \bibinfo{author}{Vaswani, A.}, \bibinfo{year}{2018}.
\newblock \bibinfo{title}{Self-attention with relative position representations}, in: \bibinfo{booktitle}{Proceedings of the 2018 Conference of the North American Chapter of the Association for Computational Linguistics: Human Language Technologies, Volume 2 (Short Papers)}, pp. \bibinfo{pages}{464--468}.
\newblock \DOIprefix\doi{10.18653/v1/n18-2074}.
\bibitem[{Shi et~al.(2023)Shi, Luo, Zhu, Kou, Cheng and Liu}]{SocialMediabenfits}
\bibinfo{author}{Shi, L.}, \bibinfo{author}{Luo, J.}, \bibinfo{author}{Zhu, C.}, \bibinfo{author}{Kou, F.}, \bibinfo{author}{Cheng, G.}, \bibinfo{author}{Liu, X.}, \bibinfo{year}{2023}.
\newblock \bibinfo{title}{A survey on cross-media search based on user intention understanding in social networks}.
\newblock \bibinfo{journal}{Information Fusion} \bibinfo{volume}{91}.
\newblock \DOIprefix\doi{10.1016/j.inffus.2022.11.017}.
\bibitem[{Shirkhorshidi et~al.(2015)Shirkhorshidi, Aghabozorgi and Wah}]{cosine_similarity}
\bibinfo{author}{Shirkhorshidi, A.S.}, \bibinfo{author}{Aghabozorgi, S.}, \bibinfo{author}{Wah, T.Y.}, \bibinfo{year}{2015}.
\newblock \bibinfo{title}{A comparison study on similarity and dissimilarity measures in clustering continuous data}.
\newblock \bibinfo{journal}{PLoS ONE} \bibinfo{volume}{10}.
\newblock \DOIprefix\doi{10.1371/journal.pone.0144059}.
\bibitem[{Sirbu et~al.(2022)Sirbu, Sosea, Caragea, Caragea and Rebedea}]{FixMatchLS}
\bibinfo{author}{Sirbu, I.}, \bibinfo{author}{Sosea, T.}, \bibinfo{author}{Caragea, C.}, \bibinfo{author}{Caragea, D.}, \bibinfo{author}{Rebedea, T.}, \bibinfo{year}{2022}.
\newblock \bibinfo{title}{Multimodal semi-supervised learning for disaster tweet classification}, in: \bibinfo{booktitle}{Proceedings of the 29th International Conference on Computational Linguistics}, pp. \bibinfo{pages}{2711--2723}.
\bibitem[{Vaswani et~al.(2017)Vaswani, Shazeer, Parmar, Uszkoreit, Jones, Gomez, Łukasz Kaiser and Polosukhin}]{self}
\bibinfo{author}{Vaswani, A.}, \bibinfo{author}{Shazeer, N.}, \bibinfo{author}{Parmar, N.}, \bibinfo{author}{Uszkoreit, J.}, \bibinfo{author}{Jones, L.}, \bibinfo{author}{Gomez, A.N.}, \bibinfo{author}{Łukasz Kaiser}, \bibinfo{author}{Polosukhin, I.}, \bibinfo{year}{2017}.
\newblock \bibinfo{title}{[transformer] attention is all you need}.
\newblock \bibinfo{journal}{Advances in Neural Information Processing Systems} \bibinfo{volume}{2017-Decem}.
\bibitem[{Wang et~al.(2023)Wang, Zhao, Qi, Liu and Shi}]{Previous_work2}
\bibinfo{author}{Wang, C.}, \bibinfo{author}{Zhao, D.}, \bibinfo{author}{Qi, X.}, \bibinfo{author}{Liu, Z.}, \bibinfo{author}{Shi, Z.}, \bibinfo{year}{2023}.
\newblock \bibinfo{title}{A hierarchical decoder architecture for multilevel fine-grained disaster detection}.
\newblock \bibinfo{journal}{IEEE Transactions on Geoscience and Remote Sensing} \bibinfo{volume}{61}.
\newblock \DOIprefix\doi{10.1109/TGRS.2023.3264811}.
\bibitem[{Wang and Sun(2023)}]{disaster}
\bibinfo{author}{Wang, S.L.}, \bibinfo{author}{Sun, B.Q.}, \bibinfo{year}{2023}.
\newblock \bibinfo{title}{Model of multi-period emergency material allocation for large-scale sudden natural disasters in humanitarian logistics: Efficiency, effectiveness and equity}.
\newblock \bibinfo{journal}{International Journal of Disaster Risk Reduction} \bibinfo{volume}{85}.
\newblock \DOIprefix\doi{10.1016/j.ijdrr.2023.103530}.
\bibitem[{Weber et~al.(2023)Weber, Papadopoulos, Lapedriza, Ofli, Imran and Torralba}]{incidents1m}
\bibinfo{author}{Weber, E.}, \bibinfo{author}{Papadopoulos, D.P.}, \bibinfo{author}{Lapedriza, A.}, \bibinfo{author}{Ofli, F.}, \bibinfo{author}{Imran, M.}, \bibinfo{author}{Torralba, A.}, \bibinfo{year}{2023}.
\newblock \bibinfo{title}{Incidents1m: A large-scale dataset of images with natural disasters, damage, and incidents}.
\newblock \bibinfo{journal}{IEEE Transactions on Pattern Analysis and Machine Intelligence} \bibinfo{volume}{45}.
\newblock \DOIprefix\doi{10.1109/TPAMI.2022.3191996}.
\bibitem[{Wei et~al.(2024)Wei, Huang, Zhao, Yu and Xu}]{constraint}
\bibinfo{author}{Wei, X.}, \bibinfo{author}{Huang, J.}, \bibinfo{author}{Zhao, R.}, \bibinfo{author}{Yu, H.}, \bibinfo{author}{Xu, Z.}, \bibinfo{year}{2024}.
\newblock \bibinfo{title}{Multi-label text classification model based on multi-level constraint augmentation and label association attention}.
\newblock \bibinfo{journal}{ACM Transactions on Asian and Low-Resource Language Information Processing} \bibinfo{volume}{23}.
\newblock \DOIprefix\doi{10.1145/3586008}.
\bibitem[{Wu et~al.(2019)Wu, Hsieh, Li and Sharpnack}]{SSE}
\bibinfo{author}{Wu, L.}, \bibinfo{author}{Hsieh, C.J.}, \bibinfo{author}{Li, S.}, \bibinfo{author}{Sharpnack, J.}, \bibinfo{year}{2019}.
\newblock \bibinfo{title}{Stochastic shared embeddings: Data-driven regularization of embedding layers}, in: \bibinfo{booktitle}{Advances in Neural Information Processing Systems}, p. \bibinfo{pages}{24–34}.
\bibitem[{Wu et~al.(2022)Wu, Mao, Xie and Li}]{Correlation}
\bibinfo{author}{Wu, X.}, \bibinfo{author}{Mao, J.}, \bibinfo{author}{Xie, H.}, \bibinfo{author}{Li, G.}, \bibinfo{year}{2022}.
\newblock \bibinfo{title}{Identifying humanitarian information for emergency response by modeling the correlation and independence between text and images}.
\newblock \bibinfo{journal}{Information Processing and Management} \bibinfo{volume}{59}.
\newblock \DOIprefix\doi{10.1016/j.ipm.2022.102977}.
\bibitem[{Xie et~al.(2022)Xie, Hou, Yu, Zhang, Luo and Zhu}]{xie}
\bibinfo{author}{Xie, S.}, \bibinfo{author}{Hou, C.}, \bibinfo{author}{Yu, H.}, \bibinfo{author}{Zhang, Z.}, \bibinfo{author}{Luo, X.}, \bibinfo{author}{Zhu, N.}, \bibinfo{year}{2022}.
\newblock \bibinfo{title}{Multi-label disaster text classification via supervised contrastive learning for social media data}.
\newblock \bibinfo{journal}{Computers and Electrical Engineering} \bibinfo{volume}{104}.
\newblock \DOIprefix\doi{10.1016/j.compeleceng.2022.108401}.
\bibitem[{Yoon et~al.(2023)Yoon, Choi and Choi}]{RoBERTaMFT}
\bibinfo{author}{Yoon, J.H.}, \bibinfo{author}{Choi, G.H.}, \bibinfo{author}{Choi, C.}, \bibinfo{year}{2023}.
\newblock \bibinfo{title}{Multimedia analysis of robustly optimized multimodal transformer based on vision and language co-learning}.
\newblock \bibinfo{journal}{Information Fusion} \bibinfo{volume}{100}.
\newblock \DOIprefix\doi{10.1016/j.inffus.2023.101922}.
\bibitem[{Zhang et~al.(2020)Zhang, Yin, Chen and Nichele}]{Emotion}
\bibinfo{author}{Zhang, J.}, \bibinfo{author}{Yin, Z.}, \bibinfo{author}{Chen, P.}, \bibinfo{author}{Nichele, S.}, \bibinfo{year}{2020}.
\newblock \bibinfo{title}{Emotion recognition using multi-modal data and machine learning techniques: A tutorial and review}.
\newblock \bibinfo{journal}{Information Fusion} \bibinfo{volume}{59}.
\newblock \DOIprefix\doi{10.1016/j.inffus.2020.01.011}.

\end{thebibliography}

\end{document}